\documentclass[10pt,conference,letterpaper]{IEEEtran}
\usepackage{times,amsmath,epsfig}
\title{Onion Curve: A Space Filling Curve with Near-Optimal Clustering}
\author{%
{Pan Xu{\small $~^{\#1}$}, Cuong Nguyen{\small $~^{*2}$}, Srikanta Tirthapura{\small $~^{*3}$} }%
\vspace{1.6mm}\\
\fontsize{10}{10}\selectfont\itshape
$^{\#}$\,University of Maryland, College Park, MD, USA\\
\fontsize{9}{9}\selectfont\ttfamily\upshape
$^{1}$\,panxu@cs.umd.edu%
\vspace{1.2mm}\\
\fontsize{10}{10}\selectfont\rmfamily\itshape
$^{*}$\,Iowa State University, Ames, IA, USA\\
\fontsize{9}{9}\selectfont\ttfamily\upshape
$^{2}$\,bnguyen@iastate.edu
$^{3}$\,snt@iastate.edu
}

\usepackage{color}
\newtheorem{definition}{Definition}
\newtheorem{theorem}{Theorem}

\newtheorem{lemma}{Lemma}

\usepackage{graphicx}
\usepackage{subcaption}

\usepackage{microtype}
\usepackage{footnote}

\usepackage{float}

\newcommand{\onion}{{\mathcal O}}
\newcommand{\HH}{{\mathcal H}}
\newcommand{\LB}{{\mathsf {LB} }}
\newcommand{\OPT}{{\mathsf {OPT} }}

\newcommand{\sumpq}{{\mathcal S}}

\newcommand{\pistart}{\pi_{s}}
\newcommand{\piend}{\pi_{e}}

\usepackage[linesnumbered, ruled, vlined]{algorithm2e}

\newcommand{\remove}[1]{}

\usepackage{hyperref}
\hypersetup{colorlinks, citecolor=red, filecolor=blue, linkcolor=blue, urlcolor=blue}

\begin{document}

\maketitle

\begin{abstract} 
Space filling curves (SFCs) are widely used in the design of indexes
for spatial and temporal data. Clustering is a key metric for an SFC,
that measures how well the curve preserves locality in moving from
higher dimensions to a single dimension. We present the {\em onion
  curve}, an SFC whose clustering performance is provably close to
optimal for cube and near-cube shaped query sets, irrespective of the
side length of the query. We show that in contrast, the clustering
performance of the widely used Hilbert curve can be far from optimal,
even for cube-shaped queries. Since the clustering performance of an
SFC is critical to the efficiency of multi-dimensional indexes based
on the SFC, the onion curve can deliver improved performance for data
structures involving multi-dimensional data.
\end{abstract}

\section{Introduction}
\label{sec:intro}
A space filling curve (SFC) is a mapping from a multi-dimensional
universe to a single dimensional universe, and is a widely used tool
in the design of data structures for multi-dimensional data. As
described in~\cite{OM84,J1990}, such a mapping allows one to apply
indexing techniques developed for single dimensional data to index
multi-dimensional data, such as in a spatial database. SFCs have been
applied to several scenarios in managing spatial and temporal data,
such as in distributed partitioning of large spatial
data~\cite{ABM16,WS93}, multi-dimensional similarity
searching~\cite{LLS16}, load balancing in parallel
simulations~\cite{LWY+17}, cryptographic transformations on spatial
data~\cite{KHC16}, to name a few.  A search for applications of space
filling curves yields more than a thousand citations on Google
Scholar, from areas such as databases and parallel computing.

In an application of an SFC to an indexing data structure, it is
usually important to have good clustering performance and locality
preservation. A query point set, such as a rectangle in the
multi-dimensional universe, should be mapped to a point set that forms
a small number of contiguous regions in a single dimension, to the
extent possible. Many SFCs have been considered in the
literature. Orenstein and Merrett~\cite{OM84} suggested the use of the
$Z$ curve, which is based on interleaving bits from different
coordinates, to support range queries.  Faloutsos~\cite{F86,F88}
suggested the use of the Gray-code curve for partial match and range
queries. The SFC that seems to have attracted the most attention is
the Hilbert curve~\cite{hilbert}. Jagadish~\cite{J1990} considered
different space filling curves, including the column-major and
row-major curves, the $Z$ curve, the Gray-code curve, and the Hilbert
curve. His experiments showed that among those considered, an ordering
based on the Hilbert curve performed the best in preserving locality.
The locality preserving property of an SFC is captured by the {\em
  clustering number}, as defined further.

\paragraph*{Model}
Let $U$ be a discrete $d$-dimensional universe of $n$ cells, of
dimensions $\sqrt[d]{n} \times \sqrt[d]{n} \ldots \times
\sqrt[d]{n}$. In this paper, we focus our analysis on the cases
$d=2,3$\footnote{Extensions to higher dimensions are possible.}.
A space filling curve (SFC) $\pi$ on $U$ is a bijective mapping $\pi: U \rightarrow \{0,1,\ldots,n-1\}$.
A {\em query} $q$ is any subset of $U$; we use $|q|$ to denote the number of cells in it. 
In this work, we are concerned with rectangular queries, which are formed by the intersection of
halfspaces. A set of cells $C \subseteq U$ is said to be a {\em
cluster} of SFC $\pi$ if the cells in $C$ are numbered consecutively
by $\pi$.

\begin{figure*}[h!]
 \begin{minipage}{\textwidth}
   \centering
   \vspace*{-1cm}
\includegraphics[width=0.5\linewidth]{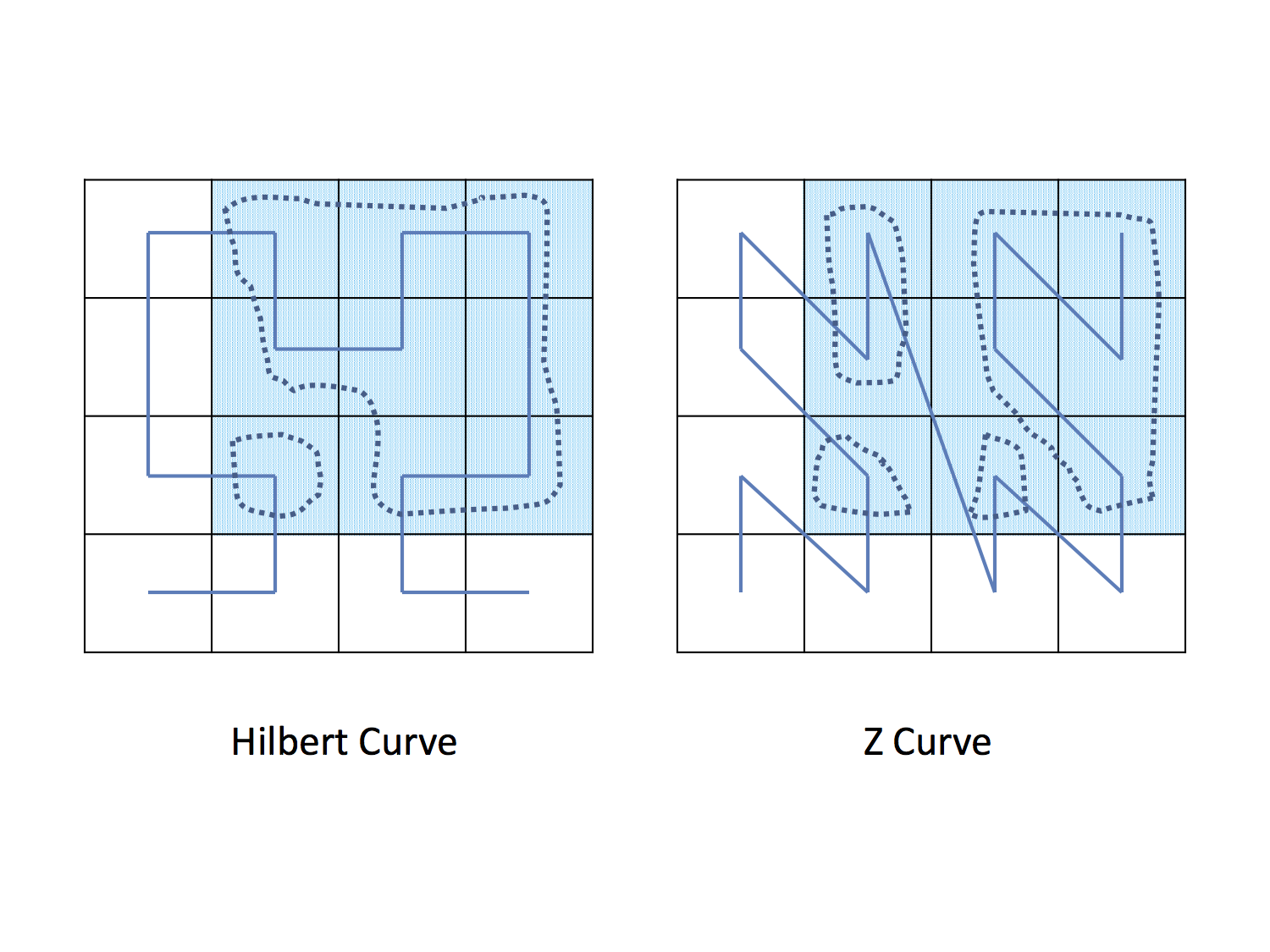}
  \vspace*{-1cm}
\caption{For the same query region shown, the Hilbert curve has a clustering number of 2, and the $Z$ curve has a clustering number of 4. Each cluster from the space filling curve is shown within a dotted region.}
\label{fig:clustering}
 \end{minipage}
\end{figure*}

The {\em clustering number} of an SFC $\pi$ for query $q$, denoted
$c(q,\pi)$, is defined as the minimum number of clusters of $\pi$ that
$q$ can be partitioned into. See Figure~\ref{fig:clustering} for an
example. The importance of the clustering number is as follows.
Suppose that multi-dimensional data was indexed on the disk according
to the ordering induced by the SFC. If the query is to retrieve all
points in a multi-dimensional region, then the clustering number
measures the number of disk ``seeks'' that need to be performed in the
retrieval. Since a disk seek is an expensive operation, a smaller
clustering number means better performance. If the query is to compute
some other function on a multi-dimensional region, then the clustering
number measures the number of sub-queries on one-dimensional ranges,
each of which can be handled efficiently using methods for indexing
one-dimensional data. Hence the general goal is to design an SFC whose
clustering number for a ``typical" query is as small as possible. We
consider the average clustering performance of an SFC for a class of
queries, as in the following definition~\cite{MJFS01,F86,J1990}.  The
clustering number of an SFC $\pi$ for a query set $Q$ is defined as:
$$
c(Q,\pi) = \frac{\sum_{q\in Q} c(q, \pi)}{|Q|}
$$

The query sets $Q$ are constructed as follows. Consider a query shape
$q$, which is a hyper-rectangle of length $\ell_i$ along dimension
$i$. We say that $q$ is a cube if $\ell_i = \ell_j$ for all $i \neq
j$. We say that $q$ is a near-cube if 
(1)~There exists $\mu$, $0 \le \mu \le 1$ such that for all $1 \le i \le d$, 
$\ell_i=\phi_i (\sqrt[d]{n})^\mu+\psi_i$ for some constants $\phi_i >0$ and
$\psi_i$; (2)~$|\ell_j-\ell_i|=o(\sqrt[d]{n})$ for each $i \neq j$.
The definition of ``near-cube'' captures the fact that 
the dependence of the side lengths of $q$ on the grid size is the same,
and also the gap between the largest and smallest sides is a lower order of the grid size. 
The second condition above is also necessary; we show a lower bound in Section~\ref{sec:lowbound2},
such that if the second condition does not hold (but the first condition does), 
then there are query sets for which there cannot exist an SFC with
a constant approximation ratio.

We primarily consider cases when $q$ is a near-cube, or a cube. We
consider query sets $Q$ that are formed by all possible translations
of such a rectangular query shape $q$. If $q$ is a cube shaped query,
we call $Q$ as ``cube query set" and if $q$ is a near-cube shaped
query, then we call $Q$ as a ``near-cube query set".

For query set $Q$, let $\OPT(Q)$ be the smallest possible clustering
number that could achieved by any SFC. For SFC $\pi$, let
$\eta(Q,\pi)= c(Q,\pi)/\OPT(Q)$ be defined as the approximation ratio
of $\pi$ for $Q$. We focus on $\eta(Q,\pi)$ as our metric for
evaluating the performance of $\pi$ for $Q$. The lower the
approximation ratio, the better is the clustering performance of $\pi$
on $Q$. 

\subsection{Our Results}

{\bf A.} We present {\bf a new SFC, the {\em onion curve}}, with the following properties.
\begin{itemize}
\item For cube query sets of any length in two dimensions, the approximation ratio of the onion curve is at most $2.32$.  For cube query sets in three dimensions, its approximation ratio is at most $3.4$. Thus we call the onion curve {\it near-optimal} for cube queries. \\

To our knowledge, this is the first SFC that has been proved to have a constant approximation ratio for cube query sets, which are the most basic query shapes. Significantly, a single curve (the onion curve) achieves near-optimal clustering for cube-shaped query sets of arbitrary sizes. \\

\item For near-cube query sets in two dimensions, the approximation ratio of the onion curve is a constant. \\
\end{itemize}

\begin{figure*}[h!]
\centering
  \begin{minipage}{0.5\textwidth}
   \centering
   \vspace*{-0.5cm}
   \includegraphics[width=0.9\textwidth]{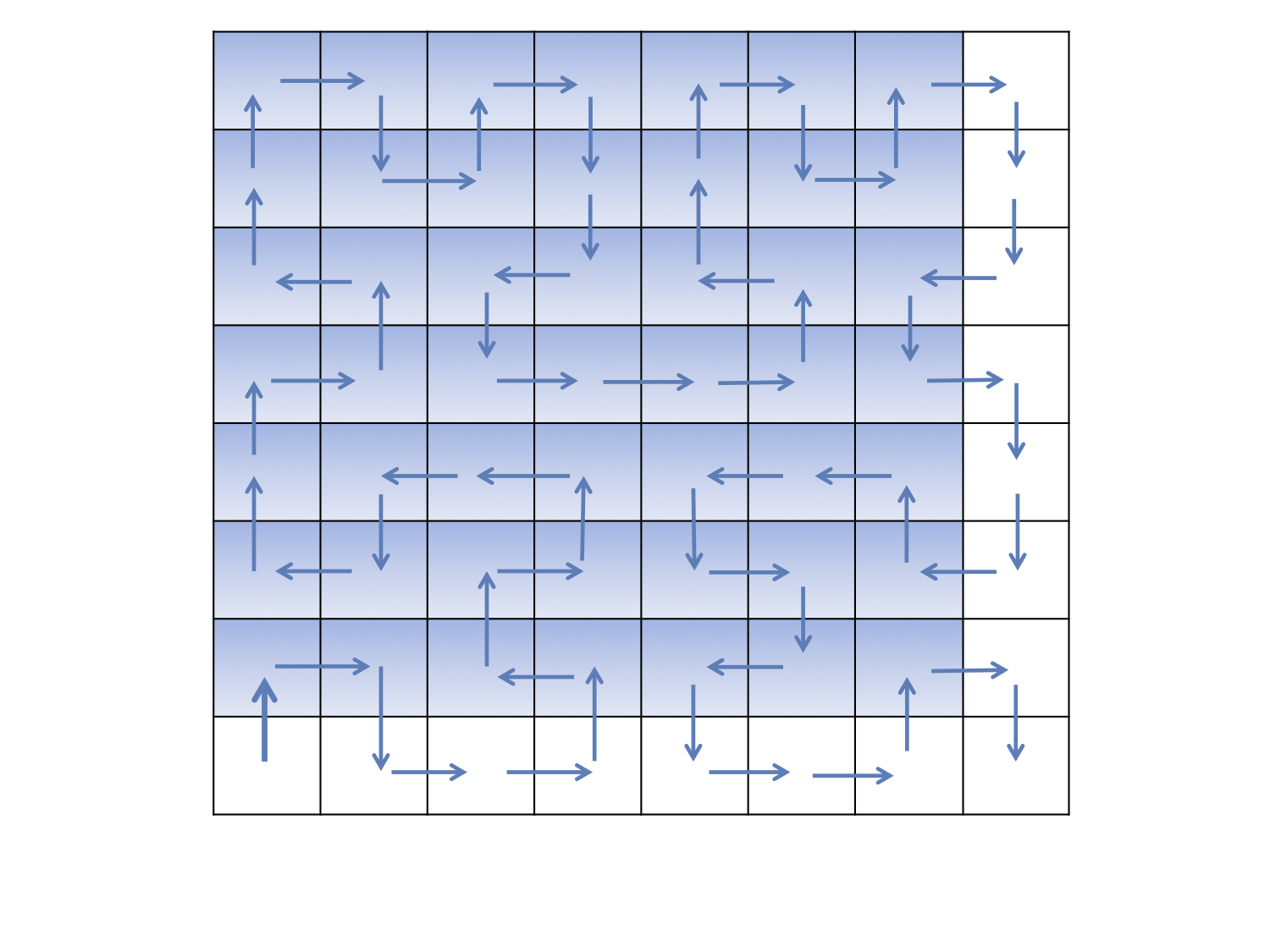}
   \vspace*{-0.3cm}
\subcaption{The Hilbert curve has a clustering number of  $5$ for the query}
\label{fig:hilbert-largequery}
    \end{minipage}\hfill
    \begin{minipage}{0.5\textwidth}
        \centering
         \includegraphics[width=0.9\textwidth]{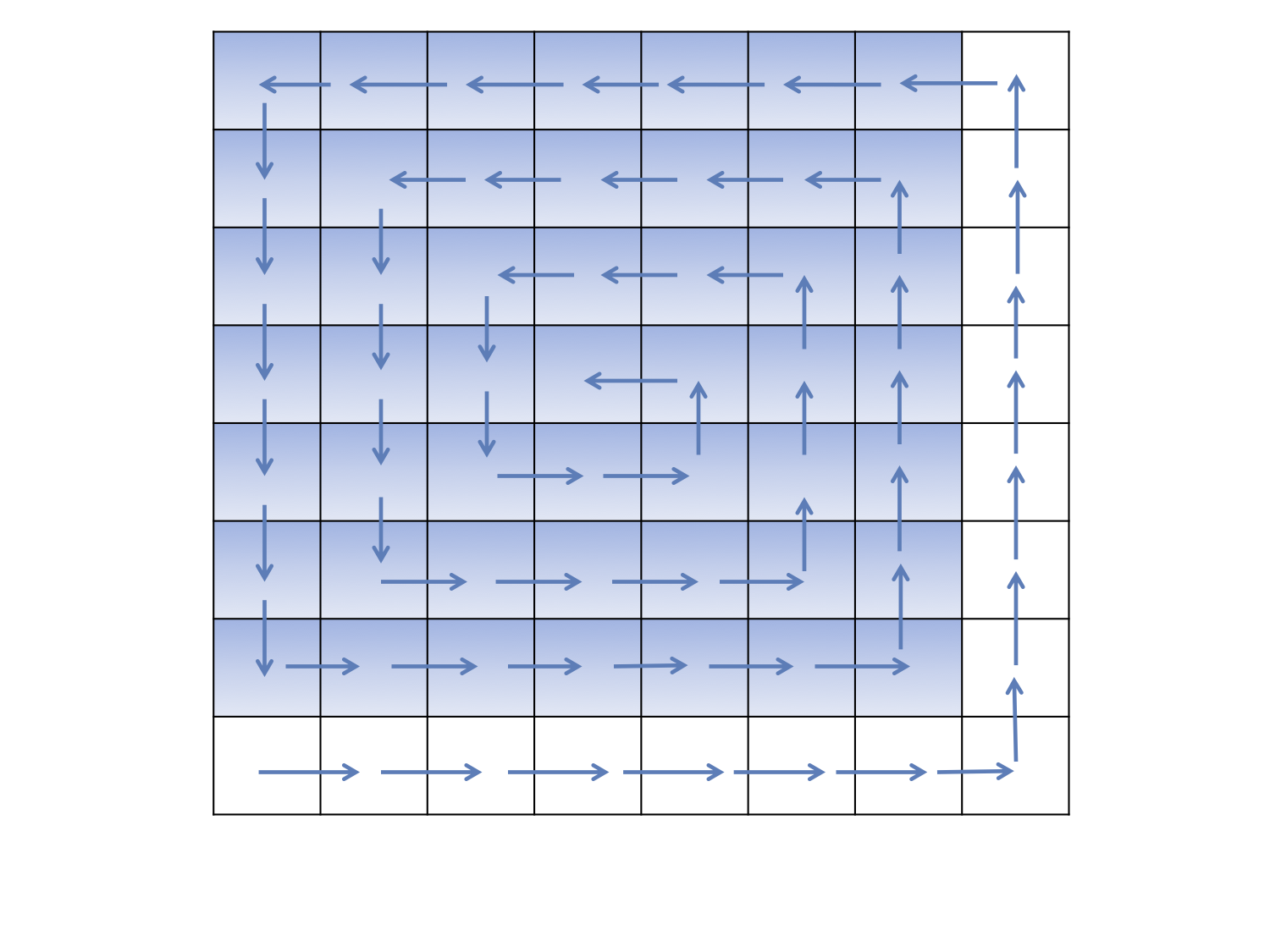}
   \vspace*{-0.5cm}
\subcaption{The onion curve has a clustering number of $1$ for the query}
\label{fig:onion-largequery}
   \end{minipage}
\caption{The average clustering number of the Hilbert curve for all $7 \times 7$ squares is much higher than that of the onion curve for the above $7 \times 7$ square query shape.}
\label{fig:hilbert-badcase}
\end{figure*}

\noindent{\bf B:} {\bf Sub-Optimality of the Hilbert curve:} In contrast, we show that the Hilbert curve can have an approximation ratio of $\Omega\left(\sqrt{n}\right)$ for cube queries in two dimensions, and an approximation ratio of $\Omega\left(n^{\frac{2}{3}}\right)$ for cube queries in three dimensions. Thus the approximation ratio of the Hilbert curve can be unbounded, even for the case of cube queries in two dimensions. See Figure~\ref{fig:hilbert-badcase} for an illustration. Detailed results are in Section~\ref{sec:hilbert}. \\

\noindent{\bf C: General Rectangular Queries:} We have shown an SFC
(the onion curve) that has near-optimal clustering (i.e. constant
approximation ratio) for sets of cube queries of any length, and also
sets of near-cube queries. It is natural to ask if it is possible to
have a single SFC that has near-optimal clustering for different
rectangular query sets. 

We show that this is impossible to achieve. There exist query sets
consisting of rectangular queries such that if an SFC performs
near-optimally on one query set, it will necessarily perform poorly on
a different one. In other words, for the class of general rectangular
queries, it is not possible to have a single SFC with a near-optimal
approximation ratio.  Details are in Lemmas~\ref{lem:row_major} and
\ref{lem:row_major_new} in Section~\ref{sec:lowbound2}.  \\



%
\noindent{\bf D:} We present results of experiments that empirically evaluate and compare the clustering performance of the onion and Hilbert curves. These results confirm our theoretical predictions and show that on cube, near-cube, and general rectangular queries, the clustering performance of the onion curve was often much better than the Hilbert curve, and always at least as good.\\

\begin{table}[ht!]
\centering
\begin{tabular}{|c|c|c|}
\hline
         & onion curve & Hilbert curve \\
\hline
Two dimensions & $2.32$  & $\Omega(\sqrt{n})$ \\
\hline
Three dimensions & $3.4$  & $\Omega(n^{\frac23})$ \\
\hline
\end{tabular}
\caption{Clustering Approximation Ratio $\eta(Q,\pi)$ for different SFCs, for cube queries
\label{tab:cube-results}}
\end{table}

The significance of this result stems from the fact that the Hilbert curve has long been considered the ``gold standard'' of SFCs for purposes of clustering. So far, we did not know of an alternative that has a demonstrably better performance. Many past analyses have almost solely focused on understanding the performance of the Hilbert curve, for example, see~\cite{MJFS01,J97}. Our results show that the onion curve consistently outperforms the Hilbert curve in terms of clustering performance. The results for cube queries are summarized in Table~\ref{tab:cube-results}.

\subsection{Related Work}
Building on earlier work~\cite{J97} that analyzed the clustering performance of the Hilbert curve on $2 \times 2$ queries, Moon {\em et al.}~\cite{MJFS01} considered an upper bound on the clustering of the Hilbert curve in $d$ dimensions.They showed that if the size of the query remains constant, the clustering number of the Hilbert curve is asymptotically equal to the surface area of the query shape divided by twice the number of dimensions. In prior work~\cite{XT14}, we generalized this result by showing that when the size of the query remains constant, the clustering number of any SFC that has the property of being ``continuous'' is asymptotically equal to that of the Hilbert curve. Further \cite{XT14} showed that for rectangular queries of constant size, the performance of the Hilbert curve is optimal. A significant difference between the analysis in prior works~\cite{XT14,MJFS01,J97} and this work is that prior works assume that the query size is constant when compared with the size of the universe, where as we do not make this assumption.

While the clustering number is an important and well-analyzed metric for the performance of an SFC, we note that there are other metrics that shed light on other properties of an SFC. In particular, prior work~\cite{GL96} has considered the ``stretch'' of an SFC, defined as the ratio between the distance between two points in the $d$-dimensional grid and the one-dimensional numbering -- such a metric is relevant to applications such as near-neighbor search. Asano {\em et al.}~\cite{ARRWW97} considered the clustering performance in a model where: for a query $q$ consisting of $|q|$ cells, the query processor is allowed to return a set of $C|q|$ cells which is a superset of $q$, and can be divided into a small number of clusters, where $C$ is a constant greater than $1$. A similar approach is considered by Haverkort~\cite{HA11}. In contrast, we require the query processor to return the set of exactly the cells present in the query $q$, and consider the number of clusters thus created. Alber and Niedermeier~\cite{AN00} present a characterization of Hilbert curve in dimensions $d \ge 3$. 

{\bf Roadmap:} The rest of this paper is organized as follows. In Section \ref{sec:gen-tec}, we present some general techniques for computing the clustering number, which are used throughout this paper. In Section~\ref{sec:2d}, we define the onion curve in two dimensions and present its analysis. In Section~\ref{sec:hilbert}, we present an analysis of the clustering number of the Hilbert curve. In Section~\ref{sec:2d-lb}, we present a lower bound on the performance of any SFC in two dimensions, allowing us to derive the approximation ratio of the onion curve. In Section~\ref{sec:3d} we present the three dimensional onion curve, its performance and related lower bounds. In Section~\ref{sec:sim}, we present results from our experimental study and conclude with a discussion in Section~\ref{sec:conclusions}.

\section{General Techniques} \label{sec:gen-tec}
In this section, we present some general techniques which will help
in computing the clustering number of an SFC and in deriving a 
lower bound on the clustering number for any SFC.

Recall that $U$ is a discrete $d$-dimensional universe of $n$ cells, of
dimensions $\sqrt[d]{n} \times \sqrt[d]{n} \ldots \times \sqrt[d]{n}$. 
An SFC $\pi$, which is a bijection from $U$ to $\{0,1,\ldots,n-1\}$,
can be viewed as a set of $(n-1)$ directed edges, 
$E(\pi) = \{(\pi^{-1}(i), \pi^{-1}(i+1)) | 0 \le i \le n-2\}$. 
Let $\pistart=\pi^{-1}(0)$ denote the first cell in $\pi$,
and $\piend=\pi^{-1}(n-1)$ denote the final cell in $\pi$.
For a directed edge $e=(\alpha,\beta)$ and query $q$, we say ``$e$
enters $q$'' if $\alpha \notin q$ and $\beta \in q$. We say ``$e$
leaves $q$'' if $\alpha \in q$ and $\beta \notin q$. We say ``$e$
crosses $q$" if either $e$ enters $q$ or if $e$ leaves $q$.
For a directed edge $e$ and query $q$, let $\gamma^+(q,e)=1$ iff 
$e$ enters $q$ and $0$ otherwise; similarly let $\gamma^-(q,e)=1$ iff 
$e$ leaves $q$ and $0$ otherwise.
Let $\gamma(q,e)=\gamma^+(q,e)+\gamma^-(q,e)$.

For a set of directed edges $E$ and query set $Q$, we define
$\gamma^+(Q,E)=\sum_{e \in E} \sum_{q\in Q} \gamma^+(q,e)$. 
Similarly we define $\gamma^-(Q,E)$ and $\gamma(Q,E)$. 
For an SFC $\pi$ and query set $Q$, we define $\gamma(Q,\pi)=\gamma( Q,E(\pi)) $. 
When the context is clear, we use $e$ to denote the set of a single edge $\{e\}$, and
$q$ to denote the set of a single query $\{q\}$.


For a cell $\alpha \in U$, query $q$ and a query set $Q$,
$$
I(q,\alpha) = \left\{
  \begin{array}{cc}
    1 & \mbox{if $\alpha \in q$}\\
    0 & \mbox{otherwise}
  \end{array}.
  \right. 
$$
Let $I(Q,\alpha)=\sum_{q \in Q} I(q,\alpha)$
Let $c(Q,\pi)$ denote the average clustering number of the SFC $\pi$
over $Q$. 
\begin{lemma}
\label{lem:LBformula}
\[
c(Q,\pi) = \frac{1}{2|Q|} \Big(\gamma(Q,\pi)+ I(Q,\pistart)+I(Q,\piend)\Big)
\]
\end{lemma}

\begin{IEEEproof}
For a query $q$ with $\pistart \notin q, \piend \notin q $, we observe that each cluster in $q$ with respect to $\pi$ 
corresponds uniquely to a pair of directed edges in $\pi$, one entering $q$ and the other
leaving $q$. Thus we have $c(q,\pi)=\frac{1}{2}\gamma(q,\pi)$, if $\pistart \notin q, \piend \notin q$.

To handle the case when $q$ includes $\pistart $ or $\piend$, we have the expression:
$c(q,\pi)=\frac{1}{2}\left( \gamma(q,\pi) + I(q,\pistart) + I(q,\piend) \right)$. Thus,
\begin{align*}
c(Q,\pi) &= \frac{1}{|Q|} \sum_{q\in Q} c(q,\pi) \\
   & =  \frac{1}{2|Q|} \sum_{q\in Q} \left( \gamma(q,\pi) + I(q,\pistart)+ I(q,\piend) \right)
\end{align*}
\end{IEEEproof}

\section{Onion Curve in Two Dimensions}
\label{sec:2d}

\subsection{Onion Curve Definition}
\label{sec:2d-def}

We first define the onion curve in two dimensions.
For integral $j \ge 1$, let $U_j$ denote the two dimensional $j \times j$ universe, 
whose coordinates along each dimension range from $0$ to $(j-1)$.
Let $\onion_j$ denote the onion curve for $U_j$.  
$\onion_j$ is defined using induction on $j, j=2,4, 6, \ldots,\sqrt{n}$.

\begin{figure*}[!ht]
\centering
\includegraphics[width=0.4\textwidth]{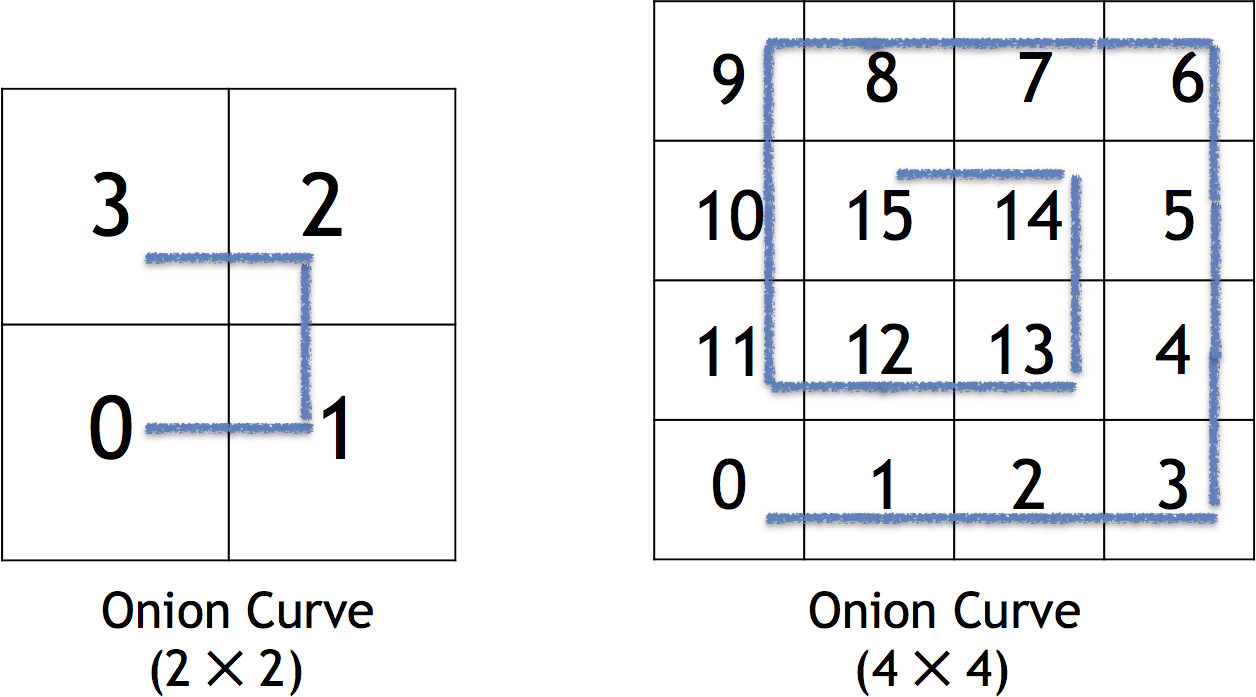}
\caption{The Two-dimensional onion Curve for the $2 \times 2$ and $4 \times 4$ universes. The number within each cell indicates the position of the cell in the onion curve.}
\label{fig:onion-new}
\end{figure*}

\begin{itemize}
\item
For $j=2$, $\onion_2$ is defined as:
$\onion_2(0,0)=0, \onion_2(1,0)=1, \onion_2(1,1)=2, \onion_2(0,1)=3$. See Figure~\ref{fig:onion-new}.

\item
For $j > 2$, $\onion_j(x_1,x_2)$ is defined as:
\begin{enumerate}
\item    $x_1$, if $x_2 = 0$
\item    $j-1 + x_2$, if $x_1 = j-1$
\item    $3j - 3 - x_1$, if $x_2 = j-1$
\item    $4j - 4 - x_2$, if $x_1 = 0, x_2 \ge 1$
\item    $4j - 4 + \onion_{j-2}(x_1-1, x_2-1)$, if $1 \le x_1,x_2 \le j-2$
\end{enumerate}
\end{itemize}

Another view of the onion curve is as follows.
For cell $\alpha = (x,y)$, let 
$\mbox{$\nabla(\alpha)=\min(x+1, \sqrt{n}-x, y+1, \sqrt{n}-y)$}$
denote the distance of the cell to the boundary of $U_{\sqrt{n}}$.  For
integer $t$, $1 \le t \le m$, let $S(t)=\{\alpha| \nabla(\alpha)=t\}$ denote the
set of cells whose distance to the boundary is $t$. $S(t)$ is also
called the ``$t$\textsuperscript{th} layer'' of the grid.
The onion curve orders all cells in $S(1)$ first, followed by cells in $S(2)$,
and so on till $S(m)$. The orders induced by the onion curve for $U_2$ and $U_4$ 
are shown in Figure~\ref{fig:onion-new}.
\vspace*{-0.1cm}
\vspace*{-0.3cm}
\subsection{Clustering Number of Onion Curve in Two Dimensions for Rectangular Queries}
\label{sec:2dub}
In this section we present an analysis of the clustering number of the two-dimensional onion curve for rectangular queries. Consider the two-dimensional universe $U$ with side length $\sqrt{n}$. 

Let the query set $Q=Q(\ell_1,\ell_2)$, equal the set formed by all possible translations of a two dimensional rectangle of lengths $\ell_1$ and $\ell_2$ respectively along dimensions $1$ and $2$. We assume $\sqrt{n}$ is even so that $m = \frac{\sqrt{n}}{2}$ is an integer. Assume without loss of generality that $\ell_1 \le \ell_2$.\footnote{Notice that the onion curve is almost symmetric to the two dimensions, i.e., it can be viewed nearly identical if we interchange the first and second dimension. Thus $c(Q(\ell_1,\ell_2),\onion)$ is almost the same as  $c(Q(\ell_2,\ell_1),\onion)$} The main result of this section is Theorem~\ref{thm:onion2d}.
Let $L_i = \sqrt{n} -\ell_i + 1, i=1,2$. 

\begin{theorem}
\label{thm:onion2d}
Let $Q=Q(\ell_1, \ell_2)$ with $\ell_1 \le \ell_2$. Let $\onion$ denote the onion curve on the $\sqrt{n} \times \sqrt{n}$ universe.
\begin{enumerate}
\item If $\ell_2 \le m$, then 
\begin{align*}
 c(Q,\onion)=&
\frac{1}{2}(\ell_1+\ell_2)+\frac{1}{L_1 L_2}\Big[ \frac{2}{3}\ell_2^3-\frac{7}{2}\ell_1 \ell_2^2+\frac{5}{2}\ell_1^2\ell_2   \\
 & - m (\ell_2-\ell_1)(\ell_2-3\ell_1) \Big]+ \epsilon_1,~~|\epsilon_1| \le 5
 \end{align*}

\item If $m<\ell_1$, 
$$c(Q,\onion)= L_1-L_2+\frac{2}{3} \frac{L_2^2}{L_1}+\epsilon_2, ~~
|\epsilon_2|\le 2$$

\end{enumerate}
\end{theorem}

{{\em Remark:} 
The results in the above two cases are
sufficient for the analysis of the Onion curve for all near-cube query
sets. From our definition, a near-cube query set $Q$ with
$\ell_1<m<\ell_2$ must have the form of $\ell_1=m+\psi_1$ and
$\ell_2=m+\psi_2$, where $\psi_1<0<\psi_2$ and both are
constants. For this case, $Q$ can be approximated by $Q'$ with
$\ell_1=\ell_2=m$ since for each given position of a two-dimensional
query $q$, expanding or reducing the query length along each dimension
by a constant will bring a difference of at most a constant in the
clustering number for the onion curve.  From the above Theorem, since $Q'$ is a cube, 
we see $c(Q', \onion) \sim \frac{2m}{3}$, which yields that $c(Q,
\onion)=\frac{2m}{3}+O(1)$. 
}

Before proving Theorem~\ref{thm:onion2d}, we will derive some useful
results that help in analyzing other two-dimensional SFCs, and in
proving lower bounds for them.

\subsubsection{Computing $\gamma(Q,\pi)$ in Two Dimensions for an SFC $\pi$} 
From Lemma \ref{lem:LBformula}, we see that for query set $Q$, computation of $\gamma(Q,\pi)$ is crucial to bounding the clustering number for an SFC $\pi$. In the following, we derive a general formula for $\gamma(Q,e)$ for the case $e=(\alpha, \beta)$ where $\alpha$ and $\beta$ are two neighboring cells on the universe, i.e., the coordinates of $\alpha$ and $\beta$ differ only $1$ in a single dimension and all others are the same.

Since $Q$ consists of all possible translations of $q$ in $U$, we can equivalently view $\gamma(Q,e)$ as the number of different positions we can translate $q$ in $U$ such that $e$ crosses $q$. Consider such a case when $e=(\alpha, \beta)$, and $\alpha$ and $\beta$ are two neighboring cells whose coordinates differ along the first dimension, but are equal along the second dimension. Observe that $e$ crosses $q$ iff (1)~either the left side or the right side of $q$ lies between $\alpha$ and $\beta$; (2) the segment between the upper and lower side of $q$ covers $e$ vertically. Notice that to achieve the first goal, we have at most $2$ options to place $q$ along the first dimension while for the second one, we have at most $\ell_2$ options along the second dimension where $\ell_2$ is the vertical side length of $q$. Formally, let $\delta_1(Q,e)$ and $\delta_2(Q,e)$, respectively equal the number of options we can place $q$ along the first and second dimension respectively such that $q$ is crossed by $e$. We present a relation between $\gamma(Q,e)$, $\delta_1(Q,e)$ and $\delta_2(Q,e)$ as follows.

For a directed edge $e=(\alpha,\beta)$ where $\alpha=(x_1, y_1)$ and $\beta=(x_2,y_2)$,
define $\nabla_1(e)=\min(x_{1}+1,\sqrt{n}-x_{1},x_{2}+1,\sqrt{n}-x_{2} )$ as
the distance of $e$ to the boundary along the first dimension. Similarly, we define
$\nabla_2(e)=\min(y_{1}+1, \sqrt{n}-y_{1},y_{2}+1, \sqrt{n}-y_{2} )$ as the distance of $e$ to the boundary along the second dimension. 


\begin{lemma}\label{lem:gamma}
Let $e=(\alpha, \beta)$ where $\alpha$ and $\beta$ are two neighboring cells along the first dimension. 
$$
\gamma(Q,e)=\delta_1(Q,e)  \cdot \delta_2(Q,e)
$$

where

\begin{itemize}
  \item
$\mbox{ if } \ell_1 \le m,
\delta_1(Q,e)=\left\{ \begin{array}{cc}
                      1 & \mbox{ if } \nabla_1(e) \le \ell_1-1  \\
                      2 &  \mbox{ if }  \nabla_1(e)  \ge \ell_1
                    \end{array}
\right\}$
  \item
  $\mbox{ if } \ell_1 > m,
\delta_1(Q,e)= \\
\left\{ \begin{array}{cc}
                      1 & \mbox{ if } \nabla_1(e) \le \sqrt{n}-\ell_1  \\
                      0 &  \mbox{ if }  \nabla_1(e)  >\sqrt{n}-\ell_1
                    \end{array}
\hspace{-0.2cm} \right\}$
  \item
$ \delta_2(Q,e)=\min\left(    \ell_2,  \sqrt{n}+1-\ell_2,    \nabla_2(e) \right) $.

\end{itemize}
\end{lemma}

\begin{IEEEproof}
Since $e=(\alpha, \beta)$ is horizontal, we have at most two choices: either the left side of $q$ or the right side of $q$ separates $\alpha$ and $\beta$. When $\ell_1 \le m$, both of the above two choices are feasible when $\nabla_1(e) \ge \ell_1$ and exactly one choice is feasible when $\nabla_1(e) \le \ell_1-1$. Similarly we can analyze the case for $\ell_1< m$. As for the feasible positions along the second dimension, all possible placements of $q$ are feasible as long as $e$ falls between the upper and lower side of $q$. Give that $e$ has the distance of $\nabla_2$ to the boundary along the second dimension and $q$ has vertical size $\ell_2$, we can verify that $\delta_2(Q,e)=\min\left(    \ell_2,  \sqrt{n}+1-\ell_2,    \nabla_2(e) \right)$. 
\end{IEEEproof}

We can get a similar result for the case when $e$ consists of two neighboring cells along the second dimension. 

\subsubsection{Clustering Number of the Onion Curve}
According to Lemma \ref{lem:LBformula}, we focus on the following sum,
in order to compute $c(Q,\onion)$:

\begin{equation}
\label{eqn:2d1}
\gamma(Q,\onion)=\sum_{e\in E(\onion)}\gamma(Q,e)
\end{equation}

Recall that $S(t)=\{\alpha| \nabla(\alpha)=t\}$ denotes the set of cells whose distance to the boundary is $t$.
We partition the edges in $E(\onion)$ into three classes,

\begin{itemize}
\item
$\mathcal{N}^1_t$: the set of edges that lie completely within a layer $S(t)$, for some
$1 \le t \le m$ while differ only along the first dimension.

\item $\widehat{\mathcal{N}^2_t}$: the set of edges that lie completely within a layer $S(t)$, for some
$1 \le t \le m$ while differ only along the second dimension. We observe from Figure~\ref{fig:onion-new} that
$\widehat{\mathcal{N}^2_t}$ is not symmetric in the way that there is one edge missing, say $e^2_t$, in the bottom of left side of $S(t)$.
Mathematically we have $e^2_t=(\alpha,\beta)$, where $\alpha=(t-1,t),\beta=(t-1,t-1)$.
For computational convenience later, we 
write $\widehat{\mathcal{N}^2_t}=\mathcal{N}^2_t-\{e^2_t\}$,
where $\mathcal{N}^2_t=\widehat{\mathcal{N}^2_t} \cup\{e^2_t\}, 1\le t \le m$.

\item
Edges that have one cell within a layer $S(t)$ and another cell
within a different layer $S(t+1)$. For each $t$, $1 \le t \le m-1$,
there is exactly one such edge in $E(\onion)$, say $e^1_t$.
\end{itemize}

Summarizing our analysis above, we have the following.
\begin{equation}\label{eqn:2d2}
\begin{split}
E(\onion) &= \left(\bigcup_{t=1}^m  \mathcal{N}^1_t \right)\bigcup  \left(\bigcup_{t=1}^m  \mathcal{N}^2_t \right) \\
& \bigcup \left\{e^1_t | 1 \le t \le m-1 \right\} -\left\{e^2_t | 1 \le t \le m \right\} 
\end{split}
\end{equation}

Observe that the first three terms on the right hand side of Equation \eqref{eqn:2d2} are three disjoint sets. Combining Equations \eqref{eqn:2d1} and \eqref{eqn:2d2},
we get:
\begin{equation}
\label{eqn:2d3}
\gamma(Q,\onion)=\sumpq_1 + \sumpq_2 + \sumpq_3
\end{equation}
where
$$\sumpq_1 =\sum_{t=1}^m \sum_{e  \in \mathcal{N}^1_t}\gamma(Q,e),~~~\sumpq_2 = \sum_{t=1}^m \sum_{e  \in \mathcal{N}^2_t}\gamma(Q,e)$$
$$\sumpq_3 =   \sum_{t=1}^{m-1} \gamma\big(Q,e^1_t\big)- \sum_{t=1}^m \gamma(Q,e^2_t)
$$

By applying Lemma \ref{lem:LBformula} and Equation~\eqref{eqn:2d3}, we have:
\begin{equation}\label{eqn:2d4}
  c(Q,\onion) =\frac{1}{2|Q|}\Big( \sumpq_1+\sumpq_2+\sumpq_3 + I(Q,\pistart)+I(Q,\piend)\Big)
\end{equation}

The following lemma shows that the contribution of $\sumpq_3$ is negligible.

\begin{lemma}\label{lem:s3less}
If $ \ell_2 \le m$, then $\frac{|\sumpq_3|}{2|Q|} \le 1$, and if $m< \ell_1$, then  $\frac{|\sumpq_3|}{2|Q|} \le \frac{1}{2}$.
\end{lemma}

\begin{IEEEproof}
We show the proof of the case $ \ell_2 \le m$. The proof of the case $m < \ell_1$ is similar.
According to Lemma \ref{lem:gamma}, we have:
$$
\sum_{t=1}^{m-1} \gamma(Q,e^1_t) \le 2(m-1) \ell_2, ~~\sum_{t=1}^m \gamma(Q,e^2_t) \le 2m \ell_1
$$

Thus we get:
$$
|\sumpq_3| =   |\sum_{t=1}^{m-1} \gamma(Q,e^1_t)- \sum_{t=1}^m \gamma(Q,e^2_t)| \le 2m \ell_2
$$
Note that $|Q|=(\sqrt{n}-\ell_1+1)(\sqrt{n}-\ell_2+1) \ge (m+1)^2$. Merging this fact with the inequality above yields our result.
\end{IEEEproof}

The next lemma offers an approximate expression for $(\sumpq_1 + \sumpq_2)$.
\begin{lemma}\label{lem:s12}
\noindent
\begin{itemize}
\item If $\ell_2 \le m$, then we have
\begin{align*}
 \frac{\sumpq_1+\sumpq_2}{2|Q|}=&
 \frac{1}{2}(\ell_1+\ell_2)
 +\frac{1}{|Q|}\Big[ \frac{2}{3}\ell_2^3-\frac{7}{2}\ell_1 \ell_2^2+\frac{5}{2}\ell_1^2\ell_2 \\
&-\frac{\sqrt{n}}{2}(\ell_2-\ell_1)(\ell_2-3\ell_1) \Big]+\epsilon_1,~ |\epsilon_1| \le 3.5
\end{align*}
 \item If $m<\ell_1$, then we have \\
$\frac{\sumpq_1+\sumpq_2}{2|Q|}= L_1-L_2+\frac{2}{3} \frac{L_2^2}{L_1}+\epsilon_2,~~~
|\epsilon_2| \le \frac{2}{3}$.
\end{itemize} 
\end{lemma}

\begin{IEEEproof}
Consider the case $\ell_2 \le m$ first. By applying Lemma~\ref{lem:gamma}, we get
$$\frac{1}{2|Q|}(\sumpq_1+\sumpq_2)=\frac{1}{2}(\ell_1+\ell_2)+\Lambda+\epsilon_1$$
where $\Lambda = \frac{1}{|Q|}\left[ \frac{2\ell_2^3}{3}-\frac{7\ell_1 \ell_2^2}{2} +\frac{5\ell_1^2\ell_2}{2} -m(\ell_2-\ell_1)(\ell_2-3\ell_1) \right]$, and $\epsilon_1 =\frac{1}{|Q|}(\frac{1}{2}-\frac{5 \ell_1^2}{2}-\frac{\ell_2}{6}+\frac{\ell_1}{2}-\frac{\ell_2^2}{2}+\frac{7}{2}\ell_2\ell_1)$. We can verify that $|\epsilon_1| \le 3.5$ when $\ell_2 \le m$. For the case when $m<\ell_1$, we have
$$\frac{1}{2|Q|}(\sumpq_1+\sumpq_2)=\frac{1}{3}(L_1+L_2)+\frac{2}{3L_1}(L_1-L_2-1)(L_1-L_2+1)$$
After expanding all the terms we get our claim for $m <\ell_1$.
\end{IEEEproof}

Now we start to prove Theorem \ref{thm:onion2d}.
\begin{IEEEproof}
Consider the case $\ell_2 \le m $ first. We have $I(Q,\pistart)+I(Q,\piend)=1+\ell_1\ell_2$ for $\pi=\onion$. Combining this with Lemma~\ref{lem:s3less}, we have
$$
\frac{1}{2|Q|}(I(Q,\pistart)+I(Q,\piend)+\sumpq_3(Q,\pistart)) \le 1.5
$$

Plugging the results above and that in Lemma \ref{lem:s12} into  Equation \ref{eqn:2d4}, we reach our conclusion. 
Similarly we can show the result in the case $m< \ell_1$.
\end{IEEEproof}

\section{Clustering Number of the Hilbert Curve}
\label{sec:hilbert}
We show that there is a sharp gap between the clustering numbers of the Hilbert curve and the onion curve, for cube queries. Figure~\ref{fig:hilbert-badcase} offers a concrete example illustrating this. Though only a single query is shown, the result is similar when we consider a query set formed by all possible translations of this query shape. Let $\mathcal{H}_{\sqrt[d]{n}}$ be a $d$-dimensional Hilbert curve filling the universe $U$ of side $\sqrt[d]{n}$. Consider the query set $Q$ formed by all possible translations of a $d$-dimensional cube query $q$ of side $\ell$, where $\ell = \sqrt[d]{n} - O(1)$. Let $L=\sqrt[d]{n}+1-\ell$, which is a constant that does not increase with $n$. The below lemma shows that the clustering number of $\mathcal{H}_{\sqrt[d]{n}}$ will be at least $\Omega\left(n^{\frac{d-1}{d}}\right)$ for $d=2,3$.

\begin{lemma} \label{lem:lb-hil}
The average clustering number of $ \mathcal{H}_{\sqrt[d]{n}}$ with respect to $Q$
is $c(Q,  \mathcal{H}_{\sqrt[d]{n}}) = \Omega\left(n^{\frac{d-1}{d}}\right)$ for $d=2,3$.
\end{lemma}

\begin{IEEEproof}
Focus on the case $d=2$. Recall that $\gamma(q,\HH)$ is total number of crossing edges of $q$ with respect to $\HH$. Let $\sqrt{n}=2^{k}$ and $U$ be the universe of side $2^k$ and $Q$ be the set of all translations of cube-query $q$ with side $\ell$. Now check the case when $\sqrt{n}=2^{k+1}$ and let $U'$ and $Q'$ be the corresponding universe and query set then. Notice that $L=\sqrt{n}+1-\ell$ is a constant, we have $|Q|=|Q'|=L^2$, which remains unchanged. Consider two queries $q \in Q$ and $q' \in Q'$ such that the relative position of $q$ to $U$  is the same as $q'$ to $U'$. In other words, the gap between $q$ and each side of $U$ is the same as that of $q'$ and $U'$. As described in~\cite{AN00}, $\mathcal{H}_{2^{k+1}}$ can be viewed as piecing together four $\mathcal{H}_{2^{k}}$ following a particular set of permutations and reflections. We can verify that $\gamma(\HH_{2^{k+1}},q') \ge 2 \gamma(\HH_{2^{k}},q)$ for each pair $(q,q')$ such that $q$ has a same relative position in $U$ as $q'$ in $U'$. 
Thus from Lemma~\ref{lem:LBformula}, we claim that $c(Q, \HH_{\sqrt{n}})$ will increase at least in the same speed as $\sqrt{n}$, which leads to our claim. The case for $d=3$ can be proved similarly. 
\end{IEEEproof}

In contrast, for the above query set, it can be seen from Theorem~\ref{thm:onion2d} that the average clustering number of the onion curve is at most $2L/3 + 2$, i.e. $\Theta(1)$. 



\section{Lower Bound in Two Dimensions}
\label{sec:2d-lb}
This section is organized as follows. In Section~\ref{sec:lowbound1},
we present a lower bound for the clustering number of a special class of SFCs that we
call {\em continuous} SFCs. We then present a lower bound for the clustering number
of any general SFC in Section~\ref{sec:lowbound2}. In Section~\ref{sec:2ddiscussion} 
we discuss the approximation ratios of the onion and Hilbert curves.

\subsection{Lower Bound for Continuous SFCs}
\label{sec:lowbound1}
For two cells $\alpha,\beta \in U$, we say $\alpha$ and $\beta$ are
neighbors iff the coordinates of $\alpha$ and $\beta$ differ by $1$
along one dimension, and are equal along the other dimension.
 
\begin{definition}
An SFC $\pi$ is said to be a {\em continuous} SFC if for any $0 \leq i \leq n-2$, $\pi^{-1}(i)$ and $\pi^{-1}(i+1)$ are neighboring cells in the universe.
\end{definition}

For example, in two dimensions, the Hilbert curve and the onion curve are continuous, while the $Z$ curve is not. 

\begin{theorem}
\label{thm:lowbound}
Let $\pi$ be any continuous SFC on $U_{\sqrt{n}}$, where $\sqrt{n}$ is even with $m=\sqrt{n}/2$.
Let $Q=Q(\ell_1, \ell_2)$ where $\ell_1 \le \ell_2$.
We have 
$
c(Q, \pi) \ge \mathsf{LB}(\ell_1, \ell_2)
$ 
where 
$$
\mathsf{LB}(\ell_1, \ell_2) =\left\{
\begin{matrix}
\frac{1}{L_1L_2}\Big(n\ell_1 + O(\sqrt{n} \ell_1 \ell_2 )\Big)  & \mbox{ if } \ell_2 \le m \\
L_2-\frac{1}{3}\frac{L_2^2}{L_1}-\epsilon,  ~0\le \epsilon \le 1 & \mbox{ if } \ell_1> m \\
\end{matrix}
\right.
$$

where $L_{i}=\sqrt{n}-\ell_i+1, i=1,2$, and 
$O(\sqrt{n} \ell_1 \ell_2 )$ is a term upper bounded by a constant factor 
of $\sqrt{n} \ell_1 \ell_2 $ when $n$ is large enough, which is detailed
as follows:

If $\ell_1 \le \ell_2/2$, $O(\sqrt{n} \ell_1 \ell_2 )=-\sqrt{n}( \ell_1 \ell_2+\frac{5}{4} \ell_1^2  )+\ell_1^2\ell_2+\frac{1}{6}\ell_1^3  +o(n \ell_1 )$.

If $\ell_1 > \ell_2/2$, $O(\sqrt{n} \ell_1 \ell_2 )=- \frac{\sqrt{n}}{4}( 9\ell_1^2+\ell_2^2 )+\frac{1}{6}\ell_1^3+3\ell_1^2\ell_2-2\ell_1\ell_2^2+\frac{1}{2}\ell_2^3+o(n \ell_1)$.
\end{theorem}



\vspace{0.1in}
\begin{definition}
\label{def:lamdaQ}
For each cell $\alpha \in U$ and query set $Q$, the minimum neighboring crossing number of $\alpha$ with respect to $Q$ is defined as:
$\lambda(Q,\alpha)=\min_{\beta \in N(\alpha) } \gamma(Q, (\alpha,\beta))$, where $N(\alpha)$ be the set of neighbors of $\alpha$ on the grid.
\end{definition}

Note that $\lambda(Q,\alpha)$ is determined exclusively by $Q$ and
$\alpha$, regardless of the SFC involved.

\begin{lemma}
\label{lem: conlower}
Consider a given continuous SFC $\pi$ and a cell $\alpha \neq \piend$. Let $e_{\pi}(\alpha)$ be the directed edge on $\pi$ starting in cell $\alpha$. We have 
$$\gamma( Q, e_{\pi}(\alpha)) \geq \lambda(Q,\alpha).$$
\end{lemma}

\begin{IEEEproof}
Since $\pi$ is continuous,
we can write $e_{\pi}(\alpha)$ as $e_{\pi}(\alpha)=(\alpha, \alpha^\prime)$ where
$\alpha^\prime \in N(\alpha)$. Thus we have $\gamma(Q,  e_{\pi}(\alpha)) \geq \lambda(Q,\alpha)$
according to Definition~\ref{def:lamdaQ}.
\end{IEEEproof}

From Lemmas \ref{lem:LBformula} and \ref{lem: conlower},
we know that the lower bound for $c(Q,\pi)$ relies primarily on the lower bound for $\sum_{\alpha \in U, \alpha \neq \piend} \lambda(Q,\alpha)$. In the following, we mainly discuss how to compute $\mathcal{T} =\sum_{i=0}^{\sqrt{n}-1} \sum_{j=0}^{\sqrt{n}-1} \lambda(i,j)$. 

Recall that for $0\leq i,j \leq \sqrt{n}-1$, $\lambda(i,j)$ is short for $\lambda(Q,\alpha)$ with $\alpha=(i,j)$. We can verify that $\lambda$ is symmetric in the following manner.
\begin{align*}
\lambda(i,j)  &= \lambda(j,i) = \lambda(i,\sqrt{n}-1-j)  \\
&= \lambda(\sqrt{n}-1-i,j) = \lambda(\sqrt{n}-1-i,\sqrt{n}-1-j)
\end{align*}

Based on symmetry, we only need to specify 
the elements of $\lambda(i,j), 0 \leq i \leq j \leq m-1 $. 
For a cell $\alpha=(i,j)$, let  $\alpha^L=(i-1,j)$ if $i \geq 1$,
$\alpha^D=(i,j-1)$ if $j \geq 1$, $\alpha^R=(i+1,j), \alpha^U=(i,j+1)$.
For each $0 \le i \le 2m-1$, let \\
$\tau(i)= \mbox{$\min(i+1, 2m-i, \ell,  2m+1-\ell )$}$.
From Lemma~\ref{lem:gamma}, we know $\lambda(i,j)$ must be achieved at  either $\gamma(Q, (\alpha, \alpha^L))$ or $\gamma(Q,(\alpha, \alpha^D))$ when $1 \le i,j \le m-1$. Directly applying Lemma~\ref{lem:gamma} to $\gamma(Q, (\alpha, \alpha^L))$ and $\gamma(Q,(\alpha, \alpha^D))$, we can get a concise formula for $\lambda(i,j)$ as follows. Let $\tau(k,\ell)=\min( k+1, \ell, 2m+1-\ell) $. Define 

$$
h_1(t, \ell)=\left\{\begin{array}{rl}
                    1 & t \leq \ell-1 \\
                    2 &  t \geq \ell
                  \end{array}
\right. 
h_2(t, \ell)=\left\{\begin{array}{rl}
                    1 & t \leq  \sqrt{n}-\ell \\
                    0 &  t \geq  \sqrt{n}-\ell+1
                  \end{array}
\right.
$$

\begin{lemma}
\label{lem:lowerBoundMatrix}
Let $ 0\leq i,j \leq m-1$. 
\begin{itemize}

\item If $\ell_2 \le m$, $\lambda(i,j)= \min \Big( h_1(i, \ell_1)\cdot\tau(j, \ell_2),  h_1(j, \ell_2)\cdot\tau(i, \ell_1)  \Big)$.
\item If $\ell_1>m$,
$\lambda(i,j)=\min\Big( h_2(i, \ell_1) \cdot\tau(j, \ell_2), h_2(j, \ell_2)\cdot \tau(i, \ell_1)  \Big) $.
\end{itemize}
\end{lemma}

Recall that $\mathcal{T}= \sum_{i=0}^{\sqrt{n}-1} \sum_{j=0}^{\sqrt{n}-1} \lambda(i,j)$. The below lemma gives an exact expression of $\mathcal{T}$.

\begin{lemma}
\mbox{}
\begin{itemize}
  \item If $  \ell_1  \le  \ell_2/2 , \ell_2 \leq m$, 
\begin{align*}
  \mathcal{T}=& 4 \Big( \frac{\ell_1}{6}- \frac{\ell_1^2}{2} + \frac{\ell_1^3}{12} - \frac{\ell_1\ell_2}{2} + \frac{\ell_1^2 \ell_2}{2} \\
  &+ \frac{3 \ell_1 m}{2}  - \frac{5 \ell_1^2 m}{4} -\ell_1 \ell_2 m + 2 \ell_1 m^2 \Big) 
\end{align*}

  \item  If $\ell_1 > \ell_2/2,  \ell_2 \leq m$, 
\begin{align*}
\mathcal{T}&=  4 \Big( \frac{\ell_1}{6}- \frac{\ell_1^2}{2} + \frac{\ell_1^3}{12} +\frac{\ell_1\ell_2}{2} + \frac{3\ell_1^2 \ell_2}{2}-\frac{\ell_2^2}{2} -\ell_1 \ell_2^2
+\frac{\ell_2^3}{4}\\
&+ \frac{ \ell_1  m}{2} - \frac{9 \ell_1^2 m}{4} +\frac{ \ell_2  m}{2}-
   \frac{ \ell_2^2 m}{4} + 2 \ell_1 m^2 \Big) 
\end{align*}

  \item If $\ell_1 >m$, $ \mathcal{T}=\frac{2}{3} (1 + 3 L_1 - L_2) L_2 (1 + L_2)$.

\end{itemize}
 \end{lemma}

Now we are ready to prove Theorem~\ref{thm:lowbound}.
\begin{IEEEproof}
For $0\leq i,j \leq \sqrt{n}-1$, we use $\lambda(i,j)$ to be short for $\lambda(Q,\alpha)$ with $\alpha=(i,j)$. Let  $\max_{0\leq i,j \leq m-1} \lambda(i,j) =\lambda_{max}$. From Lemma \ref{lem:LBformula}, we have 
\begin{align*}
 c(Q, \pi)
& =\frac{1}{2|Q|} \left[\sum_{\alpha \in U, \alpha \neq
\piend}\gamma(Q, e_{\pi}(\alpha))+
I(Q,\pistart)+I(Q,\piend) \right] \\
&\geq \frac{1}{2|Q|} \left[  \sum_{i=0}^{\sqrt{n}-1} \sum_{j=0}^{\sqrt{n}-1} \lambda(i,j) -\lambda_{max} \right]
= \frac{\mathcal{T}}{2|Q|}-\epsilon 
          \end{align*}
where $\epsilon=\frac{\lambda_{max}}{2|Q|} $. Notice that 
$\lambda_{max} \le 2 \min (L_2, L_1)$ and $|Q|=L_1L_2$, from which we see $\epsilon \le 1$. Set $\mathsf{LB}(\ell_1, \ell_2)=\frac{\mathcal{T}}{2|Q|}$ then we get our claim, where $\mathcal{T}= \sum_{i=0}^{\sqrt{n}-1} \sum_{j=0}^{\sqrt{n}-1} \lambda(i,j)$. 

\end{IEEEproof}

\subsection{Lower Bound for General SFCs}
\label{sec:lowbound2}
When $\pi$ can be an arbitrary two-dimensional SFC, we present a lower bound which is approximately half of that in the continuous case. 


\begin{theorem}
\label{thm:nonlowbound}
Let $\pi$ be an arbitrary two-dimensional SFC filling the universe
$U_{\sqrt{n}}$. Then we have
 $$c(Q, \pi) \ge \frac{1}{2}\mathsf{LB}(\ell_1, \ell_2)$$ 
 where 
 $\mathsf{LB}(\ell_1, \ell_2)$ is the exact expression as shown in Theorem~\ref{thm:lowbound}.
\end{theorem}

\begin{definition}\label{def:gen-lb}
For each cell $\alpha \in U$ and query set $Q$, the minimum crossing
number of $\alpha$ with respect to $Q$ is defined as: 
$$
\omega(Q,\alpha)=\min_{\beta \in U } \gamma(Q, (\alpha,\beta))
$$
\end{definition}

\begin{lemma}
\label{lem:noncon}
For each $\alpha \in U$, we have: $\omega(Q,\alpha) \geq  \frac{1}{2} \lambda(Q,\alpha)$.
\end{lemma}
\begin{IEEEproof}
Let $\beta$ be an arbitrary cell such that $\beta\in U, \beta \notin N(\alpha) $.
Since $\beta$ is not in $N(\alpha)$, we get that either (1) $\beta$ differs from $\alpha$ at least by $2$ in one dimension or (2) $\beta$ differs from $\alpha $ by $1$ in both of the two dimensions.
Here we focus on the first case and w.l.o.g assume $x(\beta) \ge x(\alpha)+2$, where $x(\beta)$ and $x(\alpha)$ is the $x$ coordinate of $\beta$ and $\alpha$ respectively. The second case can be analyzed in the same way. Let $\alpha^R \in N(\alpha)$ be the right neighbor of $\alpha$. We try to show that
$\gamma(Q,(\alpha,\beta)) \geq \frac{1}{2} \gamma(Q,(\alpha,\alpha^R))$.

For two cells $a$ and $b$, let $\mathcal{P}(a,b) \subseteq Q$ be the set of queries $q$ such that $a \in q, b\notin q$.
According the definition of $\gamma(Q,(\alpha,\beta))$, we have:
 $\gamma(Q,(\alpha,\beta))=|\mathcal{P}(\alpha, \beta)|+|\mathcal{P}(\beta, \alpha)|$.
Consider the following two cases respectively.
\begin{itemize}
\item $\mathcal{P}(\alpha, \alpha^R)=\emptyset$. In this case, we have 
  $ \gamma(Q,(\alpha,\alpha^R))=|\mathcal{P}(\alpha^R, \alpha)|$. For each $q\in \mathcal{P}(\alpha^R, \alpha)$, we have either  $\beta \in q$, which implies that $q \in \mathcal{P}(\beta, \alpha)$ or $q^\prime  \in (\mathcal{P}_Q(\alpha)-\mathcal{P}_Q(\beta))$ where $q^\prime$ is one unit left translation of $q$. Thus we have
\begin{align*}
\gamma(Q,(\alpha,\alpha^R)) &= |\mathcal{P}(\alpha^R, \alpha)| 
                         \le   |\mathcal{P}(\alpha, \beta)|+|\mathcal{P}(\beta, \alpha)| \\ 
                         & =\gamma(Q,(\alpha,\beta)) 
\end{align*}

\item $\mathcal{P}(\alpha, \alpha^R)\neq \emptyset$.
Notice that $\gamma(Q,(\alpha,\alpha^R)) \leq 2|\mathcal{P}_Q(\alpha,\alpha^R)|$.
One useful observation is $\mathcal{P}(\alpha,\alpha^R)\subseteq \mathcal{P}(\alpha,\beta)$. Thus we have:
\begin{align*}
 &\textstyle
\gamma(Q,(\alpha,\beta)) \ge |\mathcal{P}(\alpha,\beta)| 
                      \ge  |\mathcal{P}(\alpha,\alpha^R)| \geq \frac{1}{2}\gamma(Q,(\alpha,\alpha^R))
\end{align*}
\end{itemize}
Summarizing the above cases, we have
$\gamma(Q,(\alpha,\beta)) \geq \frac{1}{2} \gamma(Q,(\alpha,\alpha^R))$.
Since $\beta$ is an arbitrary cell such that $\beta \notin N(\alpha)$, we prove our claim.
\end{IEEEproof}

The proof of Theorem~\ref{thm:nonlowbound} follows in a very similar manner to the proof of Theorem \ref{thm:lowbound}, the main difference being that we use Lemma \ref{lem:noncon} instead of Lemma \ref{lem: conlower}.

\subsection{General Rectangular Query Sets}
Finally we use the techniques here to prove a result for the case of general rectangular queries. Consider the two dimensional query sets $Q_R$ and $Q_C$ each consisting of $\sqrt{n}$ queries, where $Q_R$ consists of all possible rows in the universe and $Q_C$ consists of all possible columns. We see that the {\em row-major} curve and {\em column-major} curve have an optimal clustering number of $1$ over $Q_R$ and $Q_C$ respectively. However,  the {\em row-major} curve performs poorly over $Q_C$ with a clustering number of $\sqrt{n}$, far from optimality; the same result for the {\em column-major} curve over $Q_R$. We now prove that {\em no SFC can have a constant clustering number over both of $Q_R$ and $Q_C$}, showing that there cannot exist a single SFC that is near-optimal for general rectangular queries.

\begin{lemma}
\label{lem:row_major}
No SFC can have a constant clustering number over both of $Q_R$ and $Q_C$, where $Q_R$ and $Q_C$ consist of all possible rows and columns in a two-dimensional universe respectively. 
\end{lemma}

\begin{IEEEproof}
Let $Q=Q_R \cup Q_C$ with $|Q|=2 \sqrt{n}$. Observe that for each $\alpha \in U$, $\omega(\alpha, Q) \ge 2$ from Definition~\ref{def:gen-lb}. By applying Lemma \ref{lem:LBformula}, we have
\begin{align*}
 c(Q, \pi) &=\frac{1}{2|Q|} \left[\sum_{\alpha \in U, \alpha \neq
\piend}\gamma(e_{\pi}(\alpha))+
I(\pistart,Q_{\ell})+I(\piend,Q_{\ell}) \right]  \\
&\ge \frac{1}{2 \sqrt{n}} \big( 2(n-1)+2 \big)=\sqrt{n}
\end{align*}
This implies that for any SFC,  the clustering number should be at least $\sqrt{n}$ over either $Q_R$ or $Q_C$.
\end{IEEEproof}

Now we give an example showing the necessity of the second condition
$\ell_2 - \ell_1 = o(n^{1/d})$ in our definition of the near-cube
query. Consider a two-dimensional query $q$ with size
$\ell_1=\sqrt{n}/2$ and $\ell_2=\sqrt{n}$ and let $q'$ be its rotation
which has $\ell_1=\sqrt{n}$ and $\ell_2=\sqrt{n}/2$.  Let $Q$ and $Q'$
be the query sets consisting of all possible translations of $q$ and
$q'$ respectively. Similar to Lemma \ref{lem:row_major}, we have

\begin{lemma}
\label{lem:row_major_new}
No SFC can have a constant clustering number over both of $Q$ and $Q'$.
\end{lemma}

Observe that (1)~both $Q$ and $Q'$ are near-cube query sets if we
remove the second condition ($\ell_2-\ell_1 = o(\sqrt{n})$) in the
definition of a near-cube query, and (2)~for either $Q$ or $Q'$, there
are SFCs (the column-major curve and the row-major curve respectively)
that achieve an optimal clustering number of $1$. Thus, no curve can
be near-optimal on $Q$ and $Q'$.


\subsection{Approximation Ratios of the Onion Curve and the Hilbert Curve for Near Cube Queries}
\label{sec:2ddiscussion}
Consider the case when $Q$ is a two-dimensional near-cube query set. Thus $Q=Q(\ell_1,\ell_2)$ where there is some constant $0 \le \mu \le 1$ such that
$\ell_i=\phi_i (\sqrt{n})^\mu+\psi_i$ for constants $\phi_i > 0$ and $\psi_i$ for $i=1,2$ and $|\ell_1-\ell_2|=o(\sqrt{n})$. Let $\eta'(Q, \pi)= c(Q, \onion)/\mathsf{LB}$
where $\mathsf{LB}$ is the lower bound for the continuous case as shown in Theorem \ref{thm:lowbound}. The approximation ratio of an SFC $\pi$ for query set $Q$ is $\eta(Q, \pi)= c(Q, \onion)/\OPT(Q)$.  
From Theorem~\ref{thm:nonlowbound}, we see that $\eta(Q,\pi) \le 2 \eta'(Q,\pi)$. Assume without loss of generality that $\ell_1 \le \ell_2$.


\begin{enumerate}
\item[I.] $\mu=0$, i.e., $\ell_1, \ell_2$ are constants that do not increase with $n$. We have $\eta'(Q, \pi)=\frac{\ell_1+\ell_2}{2 \ell_1}$. 
Note that the onion curve is continuous and almost symmetric along the two dimensions. 
Thus the results of \cite{XT12} apply when $\ell_1, \ell_2$ are constants. 
We conclude that the onion curve is optimal among all SFCs, i.e., $\eta(Q, \onion)=1$. 

\item[II.] $0<\mu<1$. $\ell_1, \ell_2$ increase at a rate slower than $\sqrt{n}$. We see that 
$$
\eta(Q, \onion) \le 2\eta'(Q, \onion) = \Big(1+\frac{\phi_2}{\phi_1}\Big)
$$

In the particular case $\phi_1=\phi_2$, we find $\eta(Q, \onion)=2$.

\item[III.] {$\mu=1$. Since $\ell_2-\ell_1=o(\sqrt{n})$, we have that $\phi_1=\phi_2$. Consider $\phi_1 = \phi_2=\phi \le 1/2$,  where $\ell_1, \ell_2$ increase at the same  rate of $\phi \sqrt{n}$, but are no more than $\sqrt{n}/2$.  In this case, we have
$$
\eta(Q, \onion) \le 2\eta'(Q, \onion) = 
2 \left(1+\frac{\phi(\frac{1}{2}-\phi)}{1-\frac{5}{2}\phi^2+\frac{5}{3}\phi^2}\right)
$$
We can verify that the rightmost expression achieves its maximum value $2.32$ when $\phi=0.355$. Thus we claim $\eta(Q, \onion) \le 2.32$. Similar results can be obtained in a slightly more general case when $\phi_1 \le \phi_2 \le 1/2$, where we can verify that $\eta(Q, \onion)$ can be expressed as a function of $\phi_1$ and $\phi_2$, even though it has two different expressions over the two cases $\phi_1 \le \phi_2/2 $ and $\phi_1 >\phi_2/2 $.



\item[IV.] $\mu=1$ and $\phi_1=\phi_2=\phi$ with $1/2<\phi<1$. We have that 
$$\eta(Q, \onion) \le 2 \eta'(Q, \onion) \le 2$$

For a slightly more general case when $ 1/2<\phi_1 \le \phi_2<1$, we have that $\eta(Q, \onion) \le 2 \eta'(Q, \onion) \le 2+3 \Big( \frac{\phi_2-\phi_1}{1-\phi_2} \Big)^2$}. 

\item[V.] $\mu=1$ and $ \phi_1 = \phi_2=1$. $\ell_1$ and $\ell_2$ differ from $\sqrt{n}$ by the constants $\psi_1  \le \psi_2 \le 0$ respectively. Notice that in this case both of $c(Q,\onion)$ and $\LB(\ell_1, \ell_2)$ are constants depending only $\psi_1$ and $\psi_2$, regardless of grid size $\sqrt{n}$. In contrast, the Hilbert curve has a clustering number of $\Omega(\sqrt{n})$, which is far from optimal. For simplifying the computation, we ignore the negligible constant terms in $c(Q, \onion)$ and $\mathsf{LB}(\ell_1,\ell_2)$.
\begin{align*}
\eta(Q, \onion) \le 2 \eta'(Q, \onion) \le 2+3 \Big( \frac{\psi_2-\psi_1}{1-\psi_2} \Big)^2
\end{align*}
Observe that in the special case $\psi_1=\psi_2$, i.e., $\ell_1=\ell_2$, we have $\eta(Q, \onion) \le 2$.
\end{enumerate}

Summarizing the five cases above, we conclude that the onion curve
achieves a constant approximation ratio for all possible near-cube
query sets.  In particular, its approximation ratio is at most $2.32$
for all cube query sets. The values of $\eta(Q,\onion)$ and $(Q,\HH)$ are summarized in
Table~\ref{tab:main-results}, for different near-cube query sets $Q$.

\begin{savenotes}
\begin{table*}[h!]
 \centering
    \begin{tabular}{| l | l | c | c | c |}
    \hline
    & \multicolumn{2}{c|}{$\eta(Q, \onion), ~~d=2$}  & $\eta(Q, \onion), d=3$ & $\eta(Q, \HH), d\in \{2,3\}$  \\
    \cline{2-2}\cline{3-3} \cline{4-4} \cline{5-5}
    & $\ell_1 \le \ell_2$ & $\ell_1=\ell_2$  & $\ell_1=\ell_2=\ell_3$&  $\ell_i=\ell, i\in [d]$ \\ \hline
$\mu=0 $ & $1$ & $1$ & $1$ & $1$ \\   \hline
$0<\mu<1$ &     $(1+\frac{\phi_2}{\phi_1})$ & $2$  & $2$ & unknown\\   \hline
$\mu=1$, $0<\phi_1 \le \phi_2 \le 1/2$ & $O(1) \footnote{It can be expressed as a function of $\phi_1$ and $\phi_2$, omitted here.}$ & $\le 2.32$ &  $ \le 3.4$ & unknown
 \\   \hline
$\mu=1$, $1/2<\phi_1 \le \phi_2 <1$ &$\le 2+3 \Big( \frac{\phi_2-\phi_1}{1-\phi_2} \Big)^2$ & 2 &  2 &unknown \\   \hline
$\mu=1$, $\phi_1 = \phi_2 =1$ & $\le  2+3 \Big( \frac{\psi_2-\psi_1}{1-\psi_2} \Big)^2$ & 2 &  $ \le 3$\footnote{We assume $\ell \le \sqrt[3]{n}-20$. We did not optimize the constant $\eta(Q,\onion)$ when $ \ell > \sqrt[3]{n}-20$.} & $\Omega(n^{\frac{d-1}{d}})$ \\ \hline
    \end{tabular}
\caption{$\eta(Q, \onion)$ and $\eta(Q,\HH)$ for different near-cube query sets $Q$. $\onion$ is the onion curve and $\HH$ is the Hilbert curve.
  \label{tab:main-results}
}
\end{table*}
\end{savenotes}

\vspace{-0.2cm}
\section{Three-dimensional Onion Curve}
\label{sec:3d}
Consider the three-dimensional universe $U$ with side length $\sqrt[3]{n}=2m$, for integer $m$. Throughout this section, we focus on the case when $Q=Q(\ell)$ which is a query set of three-dimensional cubes with side length $\ell$. The definition of onion curve for $d=3$ is similar to that of $d=2$: it numbers the cells in each layer of the universe sequentially.


\vspace{-0.2cm}
\subsection{Definition of the Onion Curve in Three Dimensions}
\label{sec:3donion}
 For a cell $\alpha = (x, y,z)$, define $\nabla(\alpha)=\min(x+1, 2m-x, y+1, 2m-y, z+1, 2m-z)$
as the distance of the cell to the boundary of $U$.  For
integer $t$, $1 \le t \le m$, let $S(t)=\{\alpha| \nabla(\alpha)=t\}$ which represents the
set of cells whose distance to the boundary is $t$. $S(t)$ is 
called the ``$t$\textsuperscript{th} layer'' of the grid. Similar to
the two-dimensional case, the onion curve numbers cells in the order $S(1)$ first, then $S(2)$, and so on till $S(m)$. For each layer $S(t), 1 \le t \le m$, we
give a disjoint partition as $S(t)=\bigcup_{g=1}^{10} S_{g}(t)$
which are defined respectively as follows. Note that for $1 \le g \le
10$, $S_g(t)$ is either a line or a two dimensional square, whose
sides are of even length.
\begin{align*}
   & \textstyle   S_1(t) =\{ i=t-1\}  \\
  & \textstyle          S_2(t) =  \{ i=2m-t\}  \\
&  \textstyle         S_3(t)= \{  t \leq i < 2m-t, j=t-1, k=t-1 \} \\
   \textstyle   &      S_4(t)=  \{  t \leq i < 2m-t,
          j=t-1,  t \leq k<2m-t \} \\
    \textstyle   &    S_5(t)=   \{
        t \leq i < 2m-t, j=t-1, k =2m-t \}  \\
     \textstyle   &    S_6(t) = \{
         t \leq i  <2m-t, j=2m-t, k =t-1 \}  \\
  \textstyle   &       S_7(t)=  \{t \leq i < 2m-t,
         j=2m-t,  t \leq k < 2m-t \}\\
  \textstyle   &      S_8(t)=     \{
         t \leq i < 2m-t, j=2m-t,   k =2m-t \} \\
  \textstyle   &      S_9(t)=        \{t \leq i < 2m-t, 
        t \leq j < 2m-t,   k =t-1 \} \\
& \textstyle           S_{10}(t)=       \{ t \leq i < 2m-t,
         t \leq j < 2m-t,   k =2m-t \}
\end{align*}

Within each layer $S(t), 1 \le t \le m$,
the onion curve will index cells in the order
$S_1(t)\rightarrow S_2(t)\rightarrow\cdots \rightarrow S_{10}(t)$.
Within each  $S_{g}(t), 1\le g\leq 10$, the onion curve then will index
each cell by the natural order induced by the line
if $S_{g}(t)$ is a line
or the order given by the two dimensional onion curve
if $S_{g}(t)$ is a two-dimensional plane.

\begin{figure*}[!ht]
\centering
\begin{minipage}{0.45\textwidth}
\centering

 \includegraphics[width=0.95\textwidth]{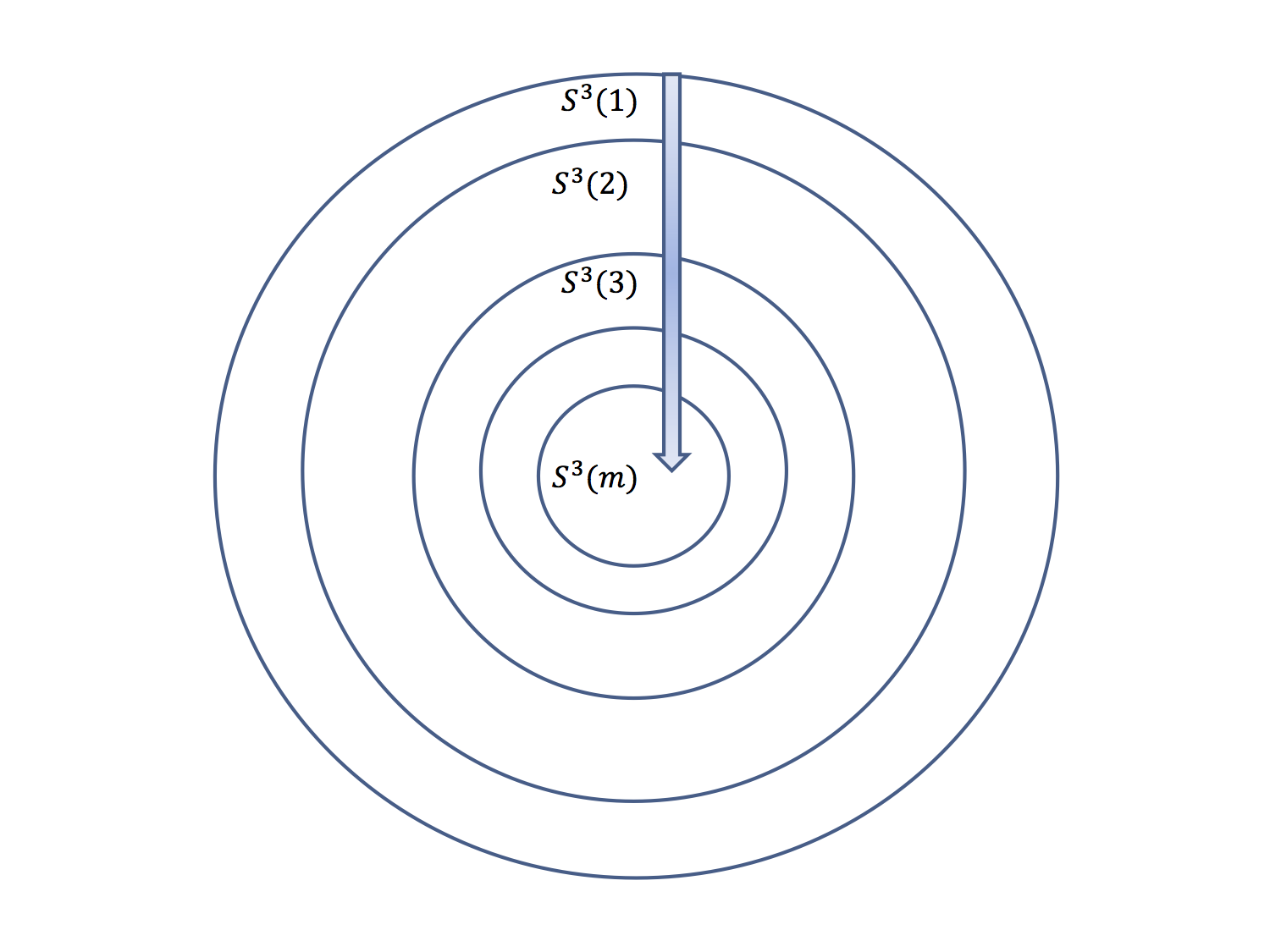}
\subcaption{The three-dimensional onion curve orders cells as $S(1), S(2),\ldots,S(t)$.}
\label{fig:3donion1}
 \end{minipage}\hfill
    \begin{minipage}{0.45\textwidth}
        \centering
        \includegraphics[width=0.95\textwidth]{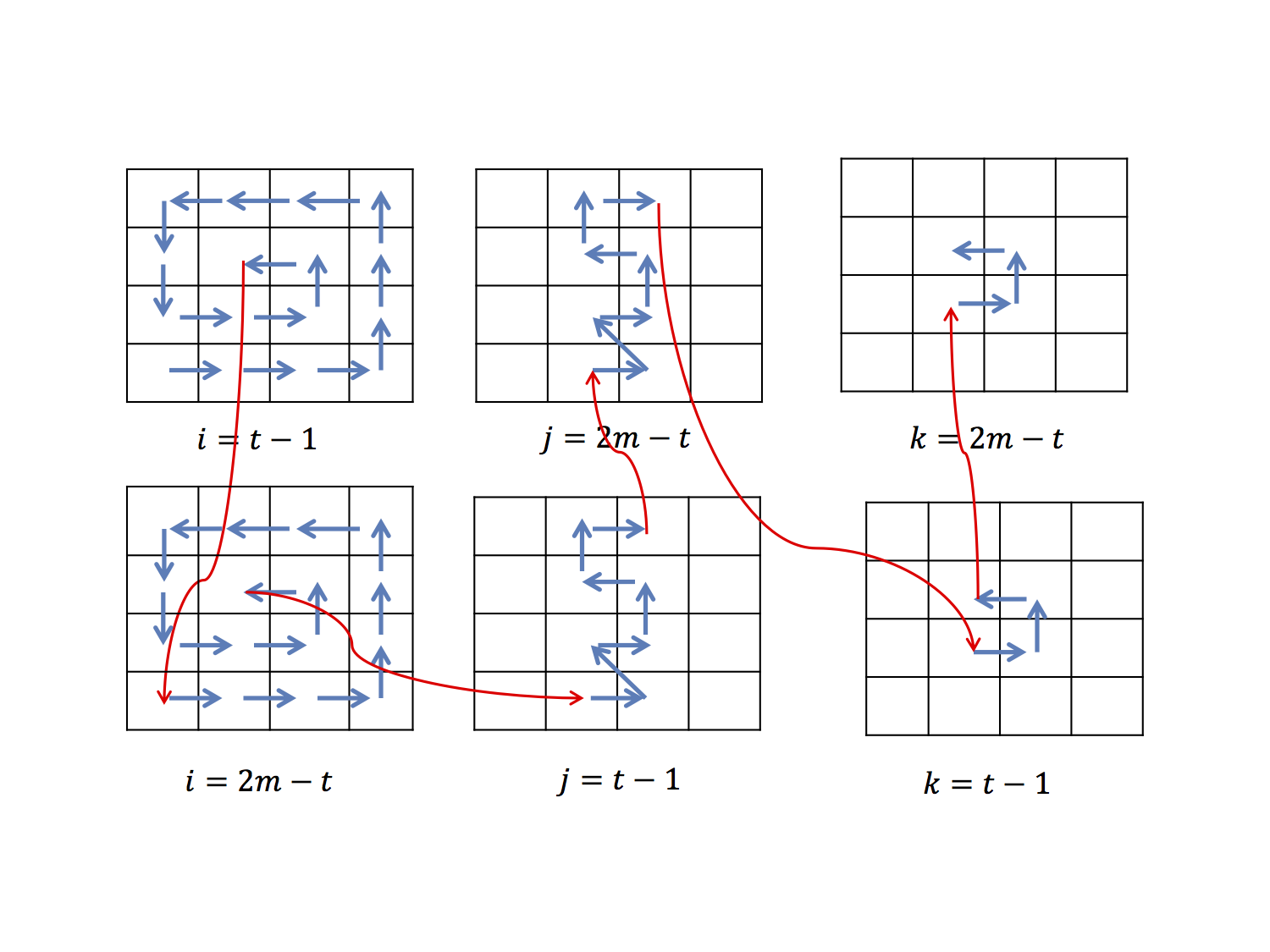}
\subcaption{The ordering among cells within $S(t), 1 \le t \le
m-1$. $i,j,k$ are respectively the coordinates along the first,
second, and third dimensions.}
\label{fig:3donion2}
\end{minipage}
\caption{Three-dimensional onion curve.}
\end{figure*}

In our definition, the essential rule that the onion curve must follow
is to organize different layers $S(t), 1\le t \le m$ sequentially
rather than intercross them.  That is the key factor that makes the
onion curve have a near-optimal clustering number, as we show later.
In contrast, the order in which the onion curve organizes the
different $S_{g}(t), 1\le g \le 10$ for each $1\le t \le m$ is not so
important. We can actually adopt any permutation on that. See
Figure~\ref{fig:3donion1} and \ref{fig:3donion2} for more details.

We now present a function, which given cell $c \in U$, returns the key of the cell, i.e. $\onion(c)$, for the three dimensional onion curve.
For a cell $\alpha$, let $t^\prime=\nabla(\alpha)$.
Let $g^\prime=g^\prime(\alpha)$ be such that $\alpha \in S_{g^\prime}(t^\prime)$.
Let $r^\prime=r^\prime(\alpha)$ be the position of $\alpha$ in $ S_{g^\prime}(t^\prime)$. From the index scheme described above, we know
 each cell $\alpha$ can be indexed by a unique triple key $(t^\prime,g^\prime,r^\prime)$. In the following, we present how to get the overall index based on its triple key.
 Let $\mathcal{K}_1(t^\prime)$ be the total number of cell in $\bigcup_{t=1}^{t^\prime-1} S(t)$.
\begin{align*}
 & \mathcal{K}_1(t^\prime)  = \sum_{t=1}^{t^\prime-1} \Big[ 2(2m-2t+2)^2 +4(2m-2t)^2+4(2m-2t) \Big]\\
& = 24m^2(t^\prime-1)-24m(t^\prime-1)^2+8(t^\prime-1)^3
\end{align*}

Let $V_{t^\prime} $ be the vector such that the
$g$th element of $V_{t^\prime}$ is $V_{t^\prime}(g)=|S_{g}(t^\prime)|, 1\le  g\leq 10$.
We can easily verify that:
\begin{eqnarray*}
  & &V_{t^\prime}(1)=V_{t^\prime}(2)=(2m-2t^\prime+2)^2;   \\
  & &V_{t^\prime}(3)=V_{t^\prime}(5)=V_{t^\prime}(6)=V_{t^\prime}(8)=2m-2t^\prime;   \\
  & & V_{t^\prime}(4)= V_{t^\prime}(7)= V_{t^\prime}(9)= V_{t^\prime}(10)=(2m-2t^\prime)^2.
\end{eqnarray*}

Let $\mathcal{K}_2(t^\prime, g^\prime)$ be the total number of cells in $\bigcup_{g=1}^{g^\prime-1} S_{g}(t^\prime)$.
i,e, $\mathcal{K}_2(t^\prime, g^\prime)=\sum_{g=1}^{g^\prime-1}V_{t^\prime}(g)$.
Finally we have that the index for cell $\alpha$ assigned by the onion curve should be:
$$
\onion(\alpha)= \mathcal{K}_1(t^\prime)+\mathcal{K}_2(t^\prime, g^\prime)+r^\prime$$

\subsection{Upper and Lower Bounds}
In this discussion, we still use $\onion$ to denote the three-dimensional onion curve. 
The following theorem gives an upper bound for $c(Q,\onion)$. All proofs are similar to the two-dimensional case and we omit the details here. 

\begin{theorem} \label{thm:3onionper}
Let $Q=Q(\ell)$ and $L=\sqrt[3]{n}-\ell+1$.
\begin{itemize} 
  \item If $\ell \leq \sqrt[3]{n}/2$, $c(Q,\onion) =\ell^2-  \frac{2}{5}\frac{\ell^5}{L^3}+o(\ell^2)$

  \item If $\ell >\sqrt[3]{n}/2$, $c(Q,\onion) \le  \frac{3}{5}L^2+\frac{13}{4}L-\frac{13}{6}$. 
\end{itemize}
\end{theorem}


The following theorem gives a lower bound on
the performance of an arbitrary continuous SFC. 
Let $L=\sqrt[3]{n}-\ell+1\ge 3$ and $m=\sqrt[3]{n}/2$.

\begin{theorem} \label{thm:3con-lowbound}
Let $\pi$ be an arbitrary {\it continuous} SFC filling the universe
$U$ and $Q=Q(\ell)$. We have $c(Q,\pi) \ge \LB(\ell)$, where:

\begin{itemize}
  \item If $2 \leq \ell \leq \sqrt[3]{n}/2$, \\
  $\LB(\ell)=\ell^2+\frac{1}{L^3}\left[\frac{29}{40}\ell^5+  \frac{15}{8}m\ell^4-3m^2 \ell^2 \right] +o(\ell^2)$,
  
    \item If $\ell >\sqrt[3]{n}/2$, $\LB(\ell) = \frac{3}{5}L^2-\frac{3}{2}L+\epsilon$ with $ 0 \le \epsilon \le 1$.
\end{itemize}
\end{theorem}

\begin{theorem} \label{thm:3lowbound}
Let $\pi$ be an arbitrary SFC filling the universe $U$ and
$Q=Q(\ell)$. We have $c(Q,\pi) \ge \frac{1}{2}\mathsf{LB}(\ell)+\epsilon$, where
$|\epsilon| \le 2$ and $\LB(\ell)$ is as shown in
Theorem~\ref{thm:3con-lowbound}.
\end{theorem}

\subsection{Discussion on the Three-dimensional Onion Curve}
\label{sec:dis_3d}
In this section, we discuss the values $\eta(Q, \onion)$ when $Q$ takes each possible cube-shaped query sets.  Similar to the two-dimensional case, we define $\eta'(Q, \onion)= c(Q, \onion)/\mathsf{LB}(\ell)$
where $\mathsf{LB}(\ell)$ is the lower bound for all {\it continuous} SFCs as shown in Theorem~\ref{thm:3con-lowbound}. From Theorem~\ref{thm:3lowbound}, we see
$\eta(Q, \onion) \le 2\eta'(Q, \onion)$. Let $\ell=\phi (\sqrt[3]{n})^\mu+\psi$ with some constants $0 \le \mu \le 1$ and $0 \le \phi$ and $\psi$.

\begin{enumerate}

\item[I.] $\mu=0$: $\ell$ is a constant. We have $ \eta'(Q, \onion)=
\frac{\ell^2}{\ell^2}=1$. Note that the three-dimensional onion curve is almost continuous and symmetric over the three dimensions. From the results in \cite{XT12}, we can further conclude that the three-dimensional onion curve is optimal among all SFCs, i.e., $\eta(Q, \onion)=1$.

  \item[II.] $0<\mu<1$: $\ell$ increases at a lower rate with 
the universe side of $\sqrt[3]{n}$. We have $ \eta(Q, \onion) \le 2$. 

\item[III.] $\mu=1$ and $0 \le \phi \le 1/2$: $\ell$ increases at the same rate of $\sqrt[3]{n}$, but is no more than $\sqrt[3]{n}/2$. 
$$\textstyle \eta(Q, \onion) \le 2\eta'(Q, \onion)=2+\frac{\frac{3}{4}\phi(\frac{1}{2}-\phi)(4+3\phi)}{(1-\phi)^3+\frac{\phi}{40}\Big(29\phi^2+\frac{75}{2}\phi -30\Big)}$$

We can verify that the rightmost expression has the maximum value of $3.4$ when $\phi=0.3967$. Thus we claim that $\eta(Q, \onion)  \le 3.4$.

\item[IV.]  $\mu=1$ and $ 1/2<\phi<1$. In this case $\ell$ increases at the same rate of $\sqrt{n}$ but large than $\sqrt[3]{n}/2$. We have $\eta(Q, \onion) \le 2$.

\item[V.]  $\mu=1$ and $\phi=1$: $\ell$ differs from $\sqrt[3]{n}$ by the constant $\psi \le 0$. In this case $L=1-\psi$ is a constant and as a result, both of $c(Q,\onion)$ and $\LB(\ell_1, \ell_2)$ are constants, regardless of grid size $\sqrt[3]{n}$. In constrast, the Hilbert curve has a clustering number of at least $\Omega(n^{\frac{2}{3}})$, which is far from optimal. For analysis convenience, we ignore the negligible constant terms included in $c(Q, \onion)$ and $\mathsf{LB}(\ell)$.

$$ \textstyle  \eta(Q, \onion)  \le 2\eta'(Q, \onion) 
      \le 2*\frac{\frac{3}{5}L^2+\frac{13}{4}L}
      {\frac{3}{5}L^2-\frac{3}{2}L}=2+\frac{95}{6}\frac{1}{(- \psi-\frac{3}{2})}$$
We can verify that $\eta(Q, \onion)  \le 3$ when $\psi \le -20$, i.e., $\ell \le \sqrt[3]{n}-20$. 
\end{enumerate}

Summarizing the discussion over the above five cases, we conclude that
the three-dimensional onion curve achieves an approximation ratio of
at most $3.4$ for nearly all cube-shaped query sets.

\section{Experimental Study}
\label{sec:sim}
\newcommand{\U}{\mathcal{U}}

We present the results of experiments that compare the performance of the onion curve and Hilbert curve over a set of cube and rectangular queries in two and three dimensions. 

\subsection{Clustering Performance on Cube Queries}
In this experiment, we consider cube queries of different lengths.

\begin{figure*}[ht!]
\centering
    \begin{minipage}{0.45\textwidth}
        \centering 
        \includegraphics[width=0.95\textwidth]{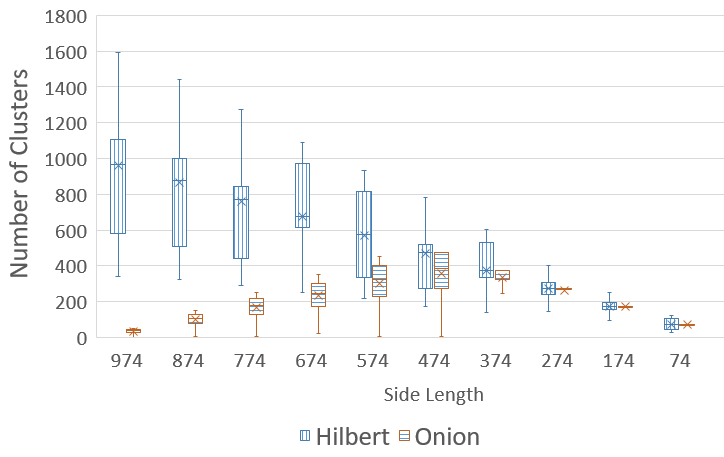}
        \subcaption{Two-dimensional Squares }
\label{fig:square2D}
\end{minipage}\hfill
\begin{minipage}{0.45\textwidth}
        \centering 
        \includegraphics[width=0.95\textwidth]{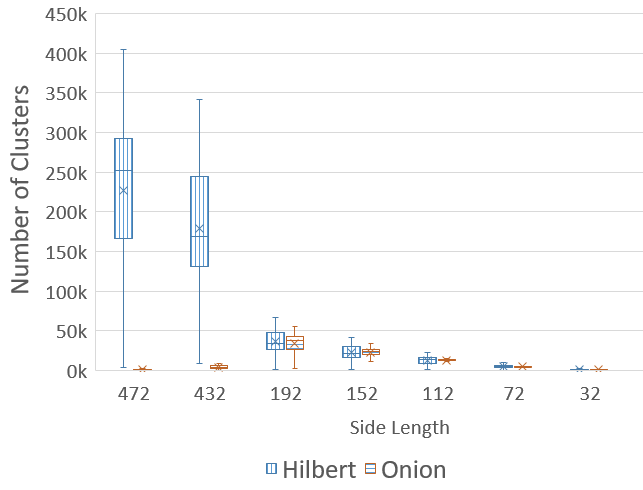}
\subcaption{Three-dimensional Cubes}
\label{fig:square3D}
\end{minipage}
\caption{The distribution of clustering numbers of the onion curve and
  Hilbert curve over a set of random squares and cubes with different
  side lengths. The box plot shows the 25 percentile and 75 percentile
  within the box, as well as the median, minimum, and maximum.}
\end{figure*}

For $d=2$, we set the length of the universe to be
$\sqrt{n}=2^{10}$. For each given $\ell=2^{10}-50\cdot k$, where 
$k \in\{1,3,5,\ldots,19\}$, we generate a set $Q(\ell)$ of $1000$ random
squares of length $\ell$. To generate a random square of a given side length, we choose the
lower left endpoint of the square uniformly among all feasible positions
for this point. For different values of $\ell$, the distribution of clustering numbers for the
two curves over $Q(\ell)$ is shown in Figure~\ref{fig:square2D}. 

Our setup for three dimensional cubes is similar.
For $d=3$, we set $\sqrt[3]{n}=2^9=512$. For each given $\ell=\{472, 432, 192, 152, 112, 72, 32\}$, we generate a
set $Q(\ell)$ of $500$ random three-dimensional cubes in the same way
as the case of $d=2$. The distribution of clustering numbers of the
two curves over each $Q(\ell)$ is shown in Figure~\ref{fig:square3D}.

From Figure~\ref{fig:square3D}, we observe the following. 
\begin{itemize}
\item
The onion curve performs much better than the Hilbert curve when the
side length of the cube is large, greater than half the length of the
axis. For instance, when $d=3$ and the side length was more than 450, 
the clustering performance of the onion curve was more than 200 times 
better than that of the Hilbert curve.

\item
For each side length considered, whether large or small,
the onion curve performed at least as well as the Hilbert curve, and
was never worse than the Hilbert curve.
\end{itemize}

\subsection{Rectangular Queries, Fixed Ratio of Side Lengths}
In this experiment, we generated a set of random rectangles where we controlled the ratio of side lengths of the rectangle. The idea is to evaluate the clustering performance of different curves on near cubes as the shape of the query moves further away from a cube. Consider the case $d=2$ first. Set $\sqrt{n}=2^{10}$. For each given point $(x,y)$ with $x \le \sqrt{n}-\ell_1, y\le  \sqrt{n}-\ell_2$, let $\mathsf{L}(x,y, \ell_1, \ell_2)$ be the rectangle with lower-left corner point as $(x,y)$ and side length as $\ell_1$ and $\ell_2$ respectively along the first and second dimension. For each given \\
$\rho \in \{\frac{1}{1024},\frac{1}{512}, \frac{1}{4},\frac{1}{2}, \frac{3}{4},1, \frac{4}{3},2,4,512,1024\}$, 
we generate a set $Q_\rho$ of random rectangles with side ratio $\rho$. The algorithm for choosing the rectangle is shown in Algorithm~\ref{alg:exp2}. 
We perform a similar experiment for the case $d=3$.

\begin{figure*}[ht!]
 \centering
    \begin{minipage}{0.45\textwidth}
        \centering
        \includegraphics[width=0.95\textwidth]{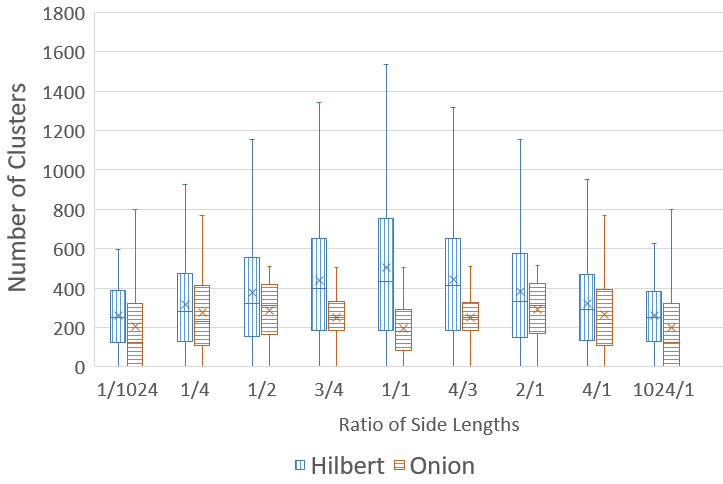}
\subcaption{Two dimensions}
\label{fig:2-ratio}
\end{minipage}\hfill
    \begin{minipage}{0.45\textwidth}
        \centering
        \includegraphics[width=0.95\textwidth]{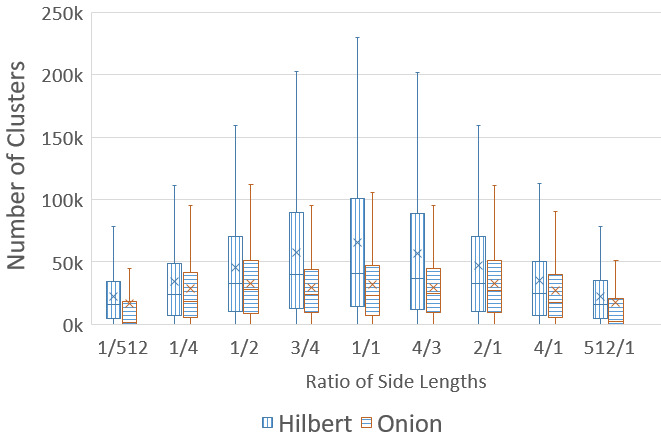}
\subcaption{Three dimensions}
\label{fig:3-ratio}
\end{minipage}
\caption{The distribution of clustering numbers of onion curve and Hilbert curve over a set of random rectangles with different ratios of side lengths in two and three dimensions.}
\end{figure*}

The box plot of distributions of clustering numbers from the onion and Hilbert curves over $Q_\rho$ is shown in Figure~\ref{fig:2-ratio} for $d=2$ and Figure~\ref{fig:3-ratio} for $d=3$ respectively. From these results, we observe that the median performance of the onion curve is better than the Hilbert curve in all cases that were considered. The difference in performance was the greatest when the ratio of side lengths approaches $1$ (i.e. the shape gets closer to a square). 

\begin{algorithm}[h!]
\caption{Generate a set $Q_\rho$ of random rectangles with a fixed side length ratio $\rho$ with $d=2$}
\label{alg:exp2}
\DontPrintSemicolon
Initialize $\sqrt{n}= \ell_2=\sqrt{n}, ~Q_\rho=\emptyset$.\;
\While{$\ell_2 \ge 0$}
{Set $\ell_1=\lfloor\ell_2/\rho \rfloor$.\;
\If{$\ell_1 \le \sqrt{n}$}
{\For{$i=1,2,\cdots,20$}{Sample $x$ and $y$ independnetly from
$\U(\sqrt{n}-\ell_1)$ and $\U(\sqrt{n}-\ell_2)$. \;
$Q_\rho \gets Q_\rho \cup\{\mathsf{L}(x,y,\ell_1,\ell_2)\}$,~~$\ell_2 \gets \ell_2-50$.
}
}
}
\end{algorithm}

\subsection{Rectangular Queries, Random End Points}
\begin{figure*}[ht!]
\centering
 \begin{minipage}{0.45\textwidth}
        \centering
        \includegraphics[width=0.95\textwidth]{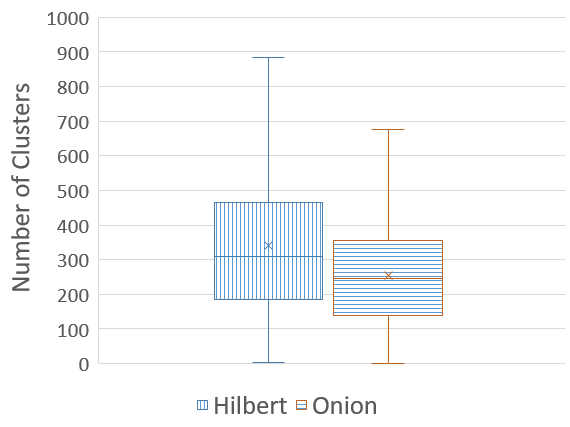}
\subcaption{Two dimensions}
\label{fig:2-random}
\end{minipage}\hfill
    \begin{minipage}{0.45\textwidth}
        \centering
        \includegraphics[width=0.95\textwidth]{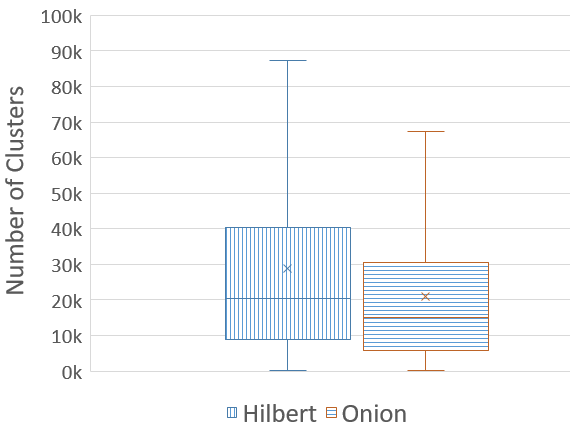}
\subcaption{Three dimensions}
\label{fig:3-random}
\end{minipage}
\caption{The distribution of clustering numbers of the onion curve and Hilbert curve over a set of random rectangles in two and three dimensions.}
\end{figure*}

In this experiment we generate a set of rectangles with random
corner points.  For an integer $N$, let $\mathcal{U}(N)$ denote the
uniform distribution over $\{0,1,2,\ldots,N\}$.  For the case $d=2$,
we generate a set $Q$ of rectangles by choosing the two corners of the
rectangle uniformly at random, and then considering the smallest
rectangle that contains both the chosen points.
We generated rectangles chosen in a similar random manner, for the case $d=3$. 
The box plot of distributions of clustering numbers from the onion and
Hilbert curves over $Q$ is shown in Figure~\ref{fig:2-random} for
$d=2$ and Figure~\ref{fig:3-random} for $d=3$ respectively.

\remove{
\begin{algorithm}[h!]
\caption{Generate a set $Q$ of rectangles with random corner points for $d=2$}
\label{alg:exp1}
\DontPrintSemicolon
Initialize $\sqrt{n}=2^{10},~~ Q=\emptyset$\;
\For{$i=1,2,\ldots, 10^4$}
{Sample $x_1, y_1, x_2, y_2$ independently from $\U(\sqrt{n})$.\;
Let $\mathsf{L}(x_1,y_1,x_2,y_2)$ be the smallest rectangle including points
$(x_1,y_1)$ and $(x_2,y_2)$. Set $Q \gets Q \cup \{\mathsf{L}(x_1,y_1,x_2,y_2)\}$
}
\end{algorithm}
}

\section{Conclusions}
\label{sec:conclusions}
We presented a new SFC, the onion curve, in two and three dimensions.
This is the first SFC we know of, whose clustering performance is
within a constant factor of the optimal for all cube and near-cube
query sets. In contrast, the clustering performance of the Hilbert
curve can be far from optimal. Further, the clustering performance of
the onion curve is no worse that the Hilbert curve in all cases
considered. The definition of the onion curve is simple, making it
easy to use.

However, we note that our analysis does not mean that the onion curve
is unambiguously better than other curves, say, the Hilbert curve, in
terms of organizing data for say range queries. There are other
aspects of clustering that we have not analyzed here, for example, the
distance between different clusters of the same query region, which
tends to be important in fetching data from the disk. Such an analysis
is interesting and the subject of future work.

The onion curve can be extended naturally to higher dimensions, using
the idea of ordering points according to increasing distance from the
edge of the universe. The analysis of such a higher dimensional onion
curve is the subject of future work. It is also interesting to
consider the performance of Hilbert curve for general near-cube query
sets as we do for the onion curve. Another future work is to complete
Table~\ref{tab:main-results} by filling the missing results for the
Hilbert curve for general near-cube query sets.

  \section{Acknowledgments}
  
The research of the first author is supported in part by NSF Awards CNS 1010789 and CCF 1422569.
\bibliographystyle{IEEEtran}

\bibliography{sfc}

\end{document}